\documentclass[preprint,aps,nofootinbib]{revtex4}
\usepackage{graphicx}
\usepackage{dcolumn}
\usepackage{bm}
\usepackage{epsf}
\usepackage{color}
\def\cE{{\cal E}}

\newcommand{\edc}{\end{document}}
\newcommand{\bb} {\begin{thebibliography}}
\newcommand{\eb}{\end{thebibliography}}
\newcommand{\bi}[1]{\bibitem{#1}}
\newcommand{\bc}{\begin{center}}
\newcommand{\ec}{\end{center}}

\newcommand{\be}{\begin{equation}}
\newcommand{\ee}{\end{equation}\normalsize}
\newcommand{\bea}{\begin{eqnarray}}
\newcommand{\eea}{\end{eqnarray}}
\newcommand{\ba}{\begin{array}{l}   }
\newcommand{\lab}[1]{\label{#1}}
\newcommand{\ea}{\end{array}}

\newcommand{\dsfrac}{\displaystyle\frac}
\newcommand{\ds} {\displaystyle}

\newcommand{\re}[1]{(\ref{#1})}

\newcommand{\ci}{\cite}

\def\bfr{{\bf r}}
\def\bfk{{\bf k}}
\def\bfq{{\bf q}}

\newcommand{{\vergul}}{  ,}

\newcommand{\veps}{\varepsilon }

\begin{document}
\draft
\title{Gr{\"u}neisen parameter of quantum magnets with spin gap}
\author{Abdulla Rakhimov$^{a}$}\email{rakhimovabd@yandex.ru}
\author{Zabardast Narzikulov$^{a}$}\email{narzikulov@inp.uz}
\author{Andreas Schilling $^{c}$}\email{schilling@physik.uzh.ch}

\affiliation{
$^a$ National University of Uzbekistan, Tashkent 100174, Uzbekistan\\
$^b$Institute of Nuclear Physics, Tashkent 100214, Uzbekistan\\
$^c$Physik-Institut, University of Z\"{u}rich, Winterthurerstrasse 190, 8057 Z\"{u}rich, Switzerland\\
}
\begin{abstract}
Using Hartree-Fock-Bogoliubov (HFB) approach 
we obtained analytical expressions for thermodynamic
quantities of the system of triplons in spin gapped quantum
magnets such as magnetization, heat capacity and the magnetic
Gr{\"u}neisen parameter $\Gamma_H$. Near the critical temperature, $\Gamma_H$
is discontinuous and changes its sign upon the Bose-Einstein
condensation (BEC) of triplons. On the other hand, in the widely
used Hartree-Fock-Popov (HFP) approach there is no discontinuity neither in the 
heat capacity nor in the  Gr{\"u}neisen parameter. We  predict that in the
low-temperature limit and near the critical magnetic field $H_c$, $\Gamma_H$ diverges
as  $\Gamma_H\sim 1/T^{2}$, while it scales as $\Gamma_H\sim 1/(H-H_c)$
as the magnetic field approaches the quantum critical point at $H_c$.

\end{abstract}
\maketitle
\section{Introduction}
The properties of condensed matter at low temperatures have always
been of high interest. Phenomena
such as novel types of superconductivity/superfluidity, quantum phase
transitions
or different types of topological order still fascinate a growing community
of researchers.
For condensed matter systems, P. Debye  and W. F.
Giauque
independently suggested in 1926 to use the magnetocaloric effect
(MCE)
of paramagnetic materials to reach temperatures significantly below 1
K. This effect,
which describes the temperature changes of a magnetic material in
response to
an adiabatic variation of the magnetic field, forms the basis of
magnetic refrigeration. The observation of a giant
MCE even around room temperature, indicating the potential of the MCE for an environment-friendly
room-temperature refrigeration, has stimulated
additional work (see recent review by Wolf \textit{et al.} \ci{wolf}).

The magnetocaloric effect (MCE) and the related magnetic Gr{\"u}neisen parameter,
\be
\lab{Grun1}
 \Gamma_{H}=\frac{1}{T}\left(\frac{\partial T}{\partial H}\right)_{S},
 \ee
quantify the cooling or heating of a
material when an applied
magnetic field is  changed under adiabatic conditions with constant
entropy $S$.
In such a process the exchanged heat is zero,
\be
 \lab{1}
 \delta Q=TdS= T\left(\frac{\partial S}{\partial T}\right)_{H}dT+
  T\left(\frac{\partial S}{\partial H}\right)_{T}dH=0,
  \ee
and hence
\be
\lab{Grun}
{\Gamma_{H}=-\frac{1}{C_{H}}\left(\frac{\partial S}{\partial
H}\right)_{T}},
 \ee
where $C_{H}=T(\partial S/\partial T)_H$ is the heat capacity at
constant magnetic field $H$.
Experimentally $\Gamma_{H}$ can be directly accessed by measuring the
change in temperature at constant entropy upon
magnetic field variation using Eq. \re{Grun1}. Mathematically
$\Gamma_{H}$ corresponds to the gradient of the temperature in the $T(H)$
landscape along an isoentropic line. Another equivalent expression for the Gr{\"u}neisen parameter
using the magnetization $M$,
\be
{\Gamma_{H}=-\frac{1}{C_H}\left(\frac{\partial M}{\partial
T}\right)_{H}},
\lab{gdmdt}
\ee
can be derived from the grand
thermodynamic potential $\Omega$ and suitable Maxwell relations using
$d\Omega=-SdT-pdV-Nd\mu-MdH$, with $\mu$ the chemical potential, $N$
the number of particles, and $p$ and $V$ pressure and volume, respectively,
which we assume in the following to be constant as we restrict
ourselves only to the magnetic subsystem.

The Gr{\"u}neisen parameter is usually discussed in terms of the quantum critical point\footnote{Below we consider only the magnetic Gr{\"u}neisen parameter}, where quantum fluctuations
play the major role. Although here we concentrate on a mean-field analysis, where these fluctuations
are not taken into account, for the sake of comparison, we briefly discuss the quantum critical point
properties.
First, if the transition occurs at a given $H_{c}$, then $\Gamma_{H}(T\to\,0,r)=G_{r}/(H-H_{c})$,
where in our notation $G_{r}\geq\,0$ is a universal prefactor \cite{Zhu,garst}.
For example, for a dilute Bose gas in the symmetry-broken state  $G_r = 1/2$ \cite{garst}.
 The temperature dependence of $\Gamma_{H}$ in the critical regime
also shows a divergence as $\Gamma_{H}(T,H\to H_{c})\sim{1}/{T^{x}}$ with a certain critical index.
It is predicted that $\Gamma_{H}$ has a different sign on each side of the quantum phase transition \cite{garst}.
These divergences and the sign change of $\Gamma_{H}$ are the hallmarks to
identify quantum critical points. These properties have been experimentally
confirmed by Gegenwart \textit{et al.},
who developed a low-frequency alternating-field
technique to measure $\Gamma_{H}$ down to low temperatures \cite{gegenw1},
in order to classify a number of magnetic systems ranging from heavy-fermion
compounds to frustrated magnets \cite{gegenw2,gegenrev}.

There is class of quantum magnets referred to as zero field
gap quantum magnets \cite{zapf,giam}.
In a subclass of these materials containing dimers of two $S=1/2$ entities, the spin gap between exited triplet ($S=1$) and
singlet ground ($S=0$) states closes beyond
a critical magnetic field $H_{c}$ due to the Zeeman effect. As a result,
bosonic quasiparticles ("triplons") arise, which may undergo a BEC below a critical
temperature $T_{c}$. Although experimental data on thermodynamic
properties are available for many of such systems (for a review, see
\ci{zapf,giam}), quantitative measurements of the MCE and the
associated Gr{\"u}neisen parameter are rare \cite{zapf,dtn2012,zvyagin}.
Experimentally it is very difficult to  explore the
behavior of $\Gamma_{H}$ in the zero-temperature limit.
This topic has not yet been systematically addressed for these materials, to the best of our knowledge,
neither theoretically nor experimentally, with, perhaps, only a single exception \cite{dtn2012}.
From simple arguments, the property  $\Gamma_{H}\sim {1}/{T^{x}}$ can
be easily considered for non-interacting Bose systems using
$C_{H}(T\to0)\sim T^{3/2}$ and $M\sim 1-{(T/T_{c})}^{3/2}$. From Eq. \re{gdmdt}, all materials
belonging to the non-interacting BEC universality
class should therefore obey $\Gamma_{H}\sim{1}/{T}$ i.e. $x=1$ \ci{dtn2012}.
Nevertheless, as the triplon bosonic quasiparticles in the magnetic
insulators to be considered here are known to
constitute an $interacting$ Bose gas \cite{zapf,nikuni,giam}, a consideration of the
effects of interaction on the Gr{\"u}neisen parameter is of utmost interest.
The aim of the present work is to investigate the properties of $\Gamma_{H}$
for such magnets within a mean field approximation.
We will show that $\Gamma_H\sim 1/(H-H_c)$ and $\Gamma_H\sim 1/T^{2}$,  and demonstrate that
$\Gamma_{H}$ changes its sign at the transition.

\section{The free energy and entropy of the triplon gas}

For $H>H_{c1}\equiv H_{c}$ the thermodynamics of a dimerized quantum magnet is determined
by the system of triplon quasiparticles with
integer spin if we neglect the phonon contribution for the moment. In
a constant external magnetic field, the number of triplons
is conserved in the thermodynamic limit, and they can experience a Bose-Einstein condensation
 (BEC)\cite{zapf,nikuni,giam}. Although the critical temperature $T_{c}$
 or the density of triplons of the BEC may be obtained within Hamiltonian
 formalism \cite{ourannals,ourprb,ournjp1,ournjp2}, the thermodynamic potential and, in particular,
 the entropy can be also derived by using a Gaussian functional
 approximation \cite{andersen}, which is in fact, equivalent to the
 Hartree-Fock-Bogoliubov (HFB) approach.

In this formalism one starts with the action
 \be
{\cal A}\left[\psi^{\dag},\psi\right]=
\int_{0}^{\beta}d\tau\int d^{3}r
 \left\{\psi^{\dag}\left[\frac{\partial}{\partial\tau}-
 \hat{K}-\mu\right]\psi+\frac{U}{2}(\psi^{\dag}\psi)^2\right\}, \label{2.1}
 \ee
 where $\beta=1/T$, $\mu$ is the chemical potential, here given as
 $\mu=\mu_{B}g(H-H_{c})$ with the Lande $g$-factor
 \cite{matsubara,zapf,nikuni,giam}, ${\hat K}$ is the operator of kinetic
 energy, $U$ represents a constant for repulsive triplon -triplon interaction, and
  $\mu_{B}$ is the Bohr magneton. The
 complex fields, $\psi^{\dag}$ and $\psi$ satisfy the standard
 bosonic periodicity conditions in that $\psi(\tau,\bfr)$ and
  $\psi^{\dag}(\tau,\bfr)$
  are periodic in $\tau$ with period $\beta$. The operator ${\hat K}$
  gives rise to the bare dispersion of triplons $\varepsilon_{k}$ as defined,
  for example,  in the bond operator representation \cite{matsumoto}. The integration in
   coordinate space may be taken in the first Brillouin zone
   with the volume $V$, which we set here $V=1$ \cite{ourlatt}. Then the thermodynamical potential $\Omega$
   can be obtained from
   \be
   \Omega=-T\ln{\cal{Z}}\lab{2.2}
   \ee
 where the grand-canonical partition function $\cal Z$ is given by the path
 integral \cite{bellac}
 \be
 {{\cal Z}=\int{\cal D}\psi^{\dag}{\cal D}\psi e^{-A[\psi^{\dag},\psi]}}.
 \lab{2.3}
 \ee
Due to the complications related to the $\psi^{4}$ term in (\ref{2.1}),
 the path integral cannot be evaluated exactly.
 In the present work we shall use a variational perturbation theory
  \cite{klbook} as outlined in Refs. \ci{ourjstatmeh,ourpra77}
for finite systems. Referring the reader to the Appendix A for the
calculation details, we obtain for $\Omega$
\bea
&&\Omega=\Omega_{\rm cl}+\Omega_{2}+\Omega_{4}, \nonumber\\
&&\Omega_{\rm cl}=-\mu\rho_{0}+\frac{U\rho_{0}^{2}}{2}+
\frac{1}{2}\sum_{k}({\cE}_k-\varepsilon_{k})+T\sum_{k}\ln(
1-e^{-\beta {\cE}_k}),  \label{8.1}\\
&&\Omega_{2}=\frac{1}{2}\left[A_{1}(U\rho_{0}-X_{2}-\mu_{1})
+A_{2}(3U\rho_{0}-X_{1}-\mu_{1})\right]\nonumber,\\
&&\Omega_{4}=\frac{U}{8}\left(3A^{2}_{1}+2A_{1}A_{2}+3A_{2}^2\right), \nonumber
\eea
and for $(i,j)$ = (1,2) or (2,1), respectively,
\be
A_{i}=G_{jj}\left.(\tau,\bfr,\tau^{\prime},\bfr^{\prime})\right
|_{\bfr\rightarrow\bfr^{\prime},
\tau\rightarrow\tau^{\prime}
}
=T\sum_{\rm n}\sum_{k}\frac{\varepsilon_{\rm k}+X_{i}}{\omega_{\rm n}^{2}+
{E}_k^{2}}=\sum_{k}\frac{\varepsilon_{k}+X_{i}}{{E}_k}W_{k},
\label{8.2}
\ee
where
\bea
&&W_{k}=\frac{1}{2}\coth\left(\frac{\beta {\cE}_k}{2}\right)=\frac{1}{2}+
n_{B}({\cE}_k)\nonumber ,\\
&&n_{B}(x)=\frac{1}{e^{x}-1},
\label{8.3}
\eea
with
\be
{\cE}_k=\sqrt{\veps_k+X_1}\sqrt{\veps_k+X_2}
\lab{bigE}
\ee
being the dispersion relation of the quasiparticles. Here $X_1$ and $X_2$ are
variational parameters defined from the principle of minimal sensitivity
\cite{andersen} as
 \be
  \ba
  \displaystyle{
  \dsfrac{\partial\Omega(X_{1},X_{2},\rho_{0})}
   {\partial X_{1}}=0},\nonumber \\
   \\
   \displaystyle{
   \dsfrac{\partial\Omega(X_{1},X_{2},\rho_{0})}
   {\partial X_{2}}=0}.
   \label{pms}
   \ea
   \ee
The normal $\rho_{1}$ and the anomalous $\sigma$ densities become
\bea
\rho_{1}=\int\langle\widetilde{\psi}^{\dag}\widetilde{\psi}\rangle d^{3}r
=\frac{A_{2}+A_{1}}{2},\nonumber\\
\sigma=\int\langle\widetilde{\psi}\widetilde{\psi}\rangle d^{3}r
=\frac{A_{2}-A_{1}}{2},
\label{8.4}
\eea
respectively. From \re{pms},  \re{8.1} and \re{8.4} one obtains for $X_{1}$ and $X_{2}$
\bea
&&X_{1}=-\mu+U\left(2\rho_1+3\rho_{0}+\sigma\right),
\label{9.1}\\
&&X_{2}=-\mu+U\left(2\rho_1+\rho_{0}-\sigma\right).\label{9.2}
\eea
The  stability condition $d\Omega/d\rho_0=0$  yields
\be
\mu-U\rho_{0}-2U\rho_{1}-U\sigma=0,\label{9.3}
\ee
where $\rho_0$ is the condensed fraction summing up to the total density $\rho=\rho_0+\rho_1$.
In general, explicit expressions for all thermodynamic quantities can be
inferred from $\Omega$ given in (\ref{8.1}). In particular,
differentiating $\Omega$ with respect to temperature yields the entropy
\be
S=-\left(\frac{\partial\Omega}{\partial T}\right)_{H}=-\sum_{k}
\ln\left[1-\exp(-\beta {{\cE}_{k}})\right]+\beta\sum_{k}
\frac{
{\cE}_{k}
}
{
(e^{\beta {\cE}_{k}}-1)
}, \label{9.4}
\ee
while the heat capacity in constant magnetic field becomes
\be
C_{H}=T\left(\frac{\partial S}{\partial T}\right)_{H}=
\beta^{2}\sum_{k}
\dsfrac
{
{\cE}_{k}({\cE}_{k}-T
 {\cE}_{k,T}^{\prime}) e^{\beta {\cE}_{k}}
 }
  {
  (e^{\beta {\cE}_{k}}-1)^2
  }.\label{9.5}
 \ee
  The resulting magnetic Gr{\"u}neisen parameter is
  \be
  \Gamma_H=-\frac{g\mu_B}{C_H}\left(\frac{\partial S}{\partial \mu}\right)_{T}=
  \frac{
  \mu_{B}g\beta^{2}
  }
  {
  C_{H}
  }
  \sum_{k}
  \dsfrac
  {
  {\cE}_{k} {{\cE}}_{k,\mu}^{\prime} e^{\beta {\cE}_{k}}
  }
 {
 (e^{\beta {\cE}_{k}}-1)^{2}
 },
 \label{9.6}
  \ee
  where ${{\cE}}_{k,T}^{\prime}=(d{\cE}_{k}/dT)_H$ and ${{\cE}}_{k,\mu}^{\prime}
  =(d{\cE}_{k}/d\mu)_{T}$, which are given explicitly in the  Appendix B.

  As we noted above, the present approximation is equivalent to the HFB
  approximation. Another similar  approach, the Hartree - Fock - Popov (HFP) approximation which is widely
   used in the literature \ci{zapf,yamada,nikuni}, can be formally obtained
   from the HFB relations by neglecting the anomalous density, i.e. by
   setting $\sigma=0$ in the above equations.

   For further considerations, we have to discuss the normal ($T\geq T_{c}$) and
   the condensed phase ($T<T_{c}$) of the system separately.

  \subsection{Normal phase, $T\geq T_{c}$}

 When the temperature exceeds a critical temperature $T\geq T_{c}$, the condensate fraction as well
 as the anomalous density vanish, i.e., $\rho_{0}=\sigma=0$, and $\rho_{1}=\rho$. In this
 normal phase both approximations, HFB and HFP, coincide.

 The basic equations (\ref{9.1}) and (\ref{9.2}) have the same
 trivial solutions as
 \be
 X_{1}=X_{2}=2U\rho-\mu.\label{10.1}
 \ee
 Inserting this into Eq. (\ref{bigE}) gives
 \be
 {E}_{k}(T\geq T_{c})\equiv\omega_{k} =\varepsilon_{k}-
  (\mu-2U\rho)\equiv\varepsilon_{k}-\mu_{\rm eff},
  \label{10.2}
  \ee
 defining the effective chemical potential $\mu_{\rm eff}$.
 Differentiating both sides of Eq. \re{10.2} with respect to $T$ and
 using Eq.  \re{9.5} gives the following expression for
 the heat capacity:
 \be
  C_H(T\geq T_c) =\beta^2\sum_{k}\dsfrac{
  \omega_k e^{\beta\omega_k}(\omega_k-2U\rho^{\prime}_T)
  }
  {(e^{\beta\omega_k}-1)^2
  }.
  \lab{cvbigtc}
  \ee

The triplon density, which defines the longitudinal magnetization
 (i.e., the component parallel to $H$) via
 \be
 M=-\frac{\partial\Omega}{\partial H}=-\frac{\partial\Omega}{\partial \mu}
 \frac{\partial \mu}{\partial H}=\mu_{B}g\rho,
 \lab{M}
 \ee
  is given by the solution of the
  nonlinear equation
  \be
  {\rho(T)=\rho_{1}=\frac{A_{1}+A_{2}}{2}=\sum_{k}\frac
  {1}{e^{\beta\omega_{k}}-1}=\sum_{k}\frac{1}
  {e^{(\varepsilon_{k}-\mu+2U\rho)\beta}-1}},\label{10.3}
  \ee
  where  we used Equations (\ref{8.2}), (\ref{8.4}) and
  (\ref{10.1}).
  Note that in this phase, the staggered magnetization $M_\perp$,
  which is a hallmark for the BEC state in dimerized spin systems, vanishes.

For the Gr{\"u}neisen parameter we have from Eqs. \re{gdmdt} and \re{M}
\be
\Gamma_H(T>T_c)=-\frac{g\mu_B}{C_H}\rho^{\prime}_T,
  \lab{grtbig}
  \ee
  where $\rho^{\prime}_T=d\rho/dT$ may be obtained from Eq. \re{10.3} (see Appendix B).

   The critical density $\rho_{c}$, i.e. the
  density of quasiparticles at the critical temperature $T_{c}$, is
  reached as soon the effective chemical potential   $\mu_{\rm eff}$ vanishes, and hence
  \be
   \rho_{c}=\rho(T_{c})=\frac{\mu}{2U}.
    \label{10.31}
     \ee
  With  this condition we may obtain the critical temperature
  as the solution of the equation
  \be
  {\frac{\mu}{2U}=\sum_{k}\frac{1}
   {
   e^{\varepsilon_{k}/T_{c}}
    -1
    }},\label{10.4}
     \ee
   which will later be used to optimize the input parameters
   of the model by comparing experimental data with the calculated $T_{c}(H)$ dependence.

   \subsection{Condensed phase, $T<T_{c}$}

In the condensed phase where the $U(1)$ symmetry is spontaneously
broken, one has to implement the Hugenholtz - Pines \cite{pines}
theorem relating the normal and the anomalous self energies
$\Sigma_{\rm n}$ and $\Sigma_{\rm an}$ to each other, i.e.
\be
\Sigma_{\rm n}-\Sigma_{\rm an}=\mu.\label{11.1}
\ee
In our notation this leads to the equation \cite{ournjp2}
\be
 X_{2}=\Sigma_{\rm n}-\Sigma_{\rm an}-\mu=0,\label{11.2}
  \ee
or
\be
 \mu-U\left(2\rho_1+\rho_{0}-\sigma\right)=0,
 \label{11.3}
  \ee
  where we have used Eq. \re{9.2}.  Due to Hugenholtz-
  Pines theorem, the excitation energy becomes gapless,
  \be
  {\cE}_k(T<T_{c})\equiv E_{k}=\sqrt{\varepsilon_{k}+X_{1}}\sqrt{\varepsilon_{k}}=
  ck+O(k^{2}),\label{11.4}
   \ee
   where $c=\sqrt{X_{1}/2m}$ is the velocity of the first sound for the
   quasiparticles with effective mass $m$. Eliminating
   $\rho_0=\rho-\rho_1$ from Eqs.
    (\ref{9.1}) and (\ref{11.3})  one  obtains the basic equation
\be
 \Delta=\frac{X_{1}}{2}=\mu+2U(\sigma-\rho_{1}),
 \label{11.5}
  \ee
  where \footnote{see ref. \cite{redbook} for the origin of the term
  $1/2$  in (\ref{11.7})}
  \bea
  \sigma&=&-\Delta\sum_{k}\frac{W_{k}}{E_{k}},\label{11.6}\\
  \rho_{1}&=&\sum_{k}\left[\frac{W_{k}(\varepsilon_{k}
   +\Delta)}{E_{k}}-\frac{1}{2}\right],
   \label{11.7}
    \eea
    and
    \be
    {E}_{k}=\sqrt{\varepsilon_{k}}\sqrt{\varepsilon_{k}
     +2\Delta}.
    \lab{dispe}
    \ee
Equation (\ref{11.3}) with $\rho_{0}=\sigma=0$
 gives the same expression for the critical density
 $\rho_{c}=\rho(T_{c})=\mu/2U$ as in Eq. \re{10.31},
 which proves  the self consistency of this approach.
Taking $d{E}_{k}/dT\equiv E^{\prime}_{k,T}$ from Eq. \re{dispe} with $E^{\prime}_{k,T}=\veps_k\Delta^{\prime}_T/E_k$
into Eq. \re{9.5} gives
\be
{C_H(T< T_c) =\beta^2\sum_{k}\dsfrac{
  e^{\beta E_k}(E_{k}^{2}-T\veps_k\Delta^{\prime}_T)
 }
 {(e^{\beta E_k}-1)^2
 }},
\lab{cvtsmc}
\ee
where $\Delta^{\prime}_T$ is given in the Appendix B.

 For practical calculations Eq. \re{11.5} can
 be rewritten as
 \be
 Z=1+\widetilde{\sigma}(Z)-\widetilde{\rho}_{1}(Z)
 \lab{Z},
 \ee
 where $Z=\Delta/\mu$, $\widetilde{\sigma}=\sigma/\rho_{c}$, and
 $\widetilde{\rho}_{1}=\rho_{1}/\rho_{c}$.
After solving the equation \re{Z}, the longitudinal and the staggered
  magnetizations $M$ and $M_\perp$ in the condensed phase,
  respectively, become
  \bea
  M(T\leq T_c)=g\mu_B\rho=g\mu_B\rho_c(Z+1), \lab{11.9}\\
  M^{2}_{\perp}(T\leq T_c)=\frac{1}{2}g^2\mu_{B}^{2}\rho_0=
  \frac{1}{2}g^2\mu_{B}^{2}\rho_c(2Z-\widetilde\sigma), \nonumber
   \eea

where we used
\be
\rho=\frac{\Delta+\mu}{2U},\quad \quad
\rho_0=\frac{\Delta}{U}-\sigma
\lab{rhotrho0}.
\ee

The Gr{\"u}neisen parameter is with Eqs. \re{11.9} and \re{rhotrho0}
\be
\Gamma_H (T\leq T_c)=-\frac{g\mu_B\rho^{\prime}_T}{C_H}=-\frac{g\mu_B\Delta^{\prime}_T}{2UC_H},
\lab{grsmtc}
\ee
where $C_H$ is given in Eq. \re{cvtsmc}.

The main difference between the HFB and HFP approximations manifests
itself in the condensed phase. In particular, the basic equation
\re{11.5} simplifies to

\be
\Delta_{\rm HFP}=\mu-2U\rho_1=U\rho_0,
\lab{hfpdelta}
\ee
where $\rho_1$ is formally the same as in Eq.  \re{11.7}.
\section{Low temperature expansion}

In the present section we will derive analytical expressions
in the $T\rightarrow 0$ limit. We shall perform the low-temperature
expansion as a function of the dimensionless parameter  $Tm=\widetilde{T}$.
In fact, for the majority of spin gap quantum magnets,
the effective mass $m$  is small, e.g. $m\approx 0.02 \mbox{ K}^{-1}$ for
TlCuCl$_3$ \ci{misgu}, so that any power series in the small parameter $\widetilde{T}$
should quickly converge.

In general, three dimensional momentum integrals e.g in Eq. \re{11.7} can not be taken
analytically. So, to overcome this difficulty we use Debye - like approximation
\ci{yuklaser}
. To this end the inegration over momentum in Brillouene zone is replaced by the Debye sphere,
 whose radius $k_D$ is chosen such that to retain the normalization condition
\be
\ba
\ds\sum_{k\in {\cal B}}\rightarrow \dsfrac{V}{(2\pi)^3}\ds\int_{{\cal B}} d{\bf{k}}
=\dsfrac{V}{(2\pi)^3}\ds\int_{-\pi/a}^{\pi/a} dk_x dk_y dk_z=
\dsfrac{1}{8}\ds\int_{-1}^{1} dq_x dq_y dq_z\approx \dsfrac{\pi}{2}
\ds\int_{0}^{Q_0} q^2 dq =1
\ea
\ee
which  gives the dimensionless Debye radius
    $Q_0=(6/\pi)^{1/3}\approx 1.24$ ,  i.e $k_D=Q_0\pi/a$.
In practical calculations we use dimensionless
momentum variable ${\bf q}={\bf k}a/\pi$.
Next, we replace  the simple symmetric three-dimensional bare dispersion
\be
\varepsilon_{k}=J_{0}(3-\cos k_{x}a-\cos k_{y}a-\cos k_{z}a),
\lab{bare}
\ee
 which is frequently used as a model dispersion relation in gapped
 quantum magnets \ci{giam}, by  $\varepsilon_{k}\approx J_{0}k^2/2\equiv k^{2}/2m$.
Then  the momentum integration
may be  approximated as
 \be
 \sum_{k}f(\varepsilon_k)=
 \frac{V}{(2\pi)^{3}}\int_{-\pi/a}
 ^{\pi/a}f(\varepsilon_{k})dk_{x}dk_{y}dk_{z}=
 \int_{0}^{1}f(\varepsilon_{\bfq})dq_{x}dq_{y}dq_{z}\approx\frac{\pi}{2}\int_{0}^
 {Q_0}q^2dqf(\varepsilon_{q})
 \ee
 where
 $\varepsilon_{q} \approx  q^2\pi^2/2m$.
   As to the phonon dispersion, similarly to the case of optical lattices,
    one may use long- wave approximation \ci{yuklaser}:
 \be
E_q =\sqrt{\veps_q}\sqrt{\veps_q+2\Delta}\approx c\pi q
\lab{3p1}
 \ee
 with the sound velocity at zero temperature $c=\sqrt{\Delta(T=0)/m}$.
With these approximations for the low-temperature limit, most of the integrals can be evaluated explicitly in terms of logarithmic and
 polylogarithmic  functions ${\rm Li}_s(z)$ of the argument
 $z=\exp(-Q_0c\pi\beta)$ , i.e., as a function  $F(T,z)$ \ci{robinson} .
 Since $z$ decreases quickly with increasing $\beta$ we may expand $F(T,z)$ in powers of $z$ to extract a leading term. On the other hand  one may also introduce the Debye temperature
 $T_D=ck_D$ and make an expansion in powers of $T/T_D$ as in solid state physics.

We refer the reader to the Appendix B for the further details of the
calculation. The final  result for the entropy becomes
\be
S=\frac{2\pi^2(\widetilde{T})^3}{45\gamma^3}  +   O(\widetilde{T}^5)
 \lab{stsm}
 \ee
where $\gamma=cm$ and  $\widetilde{T}=Tm$. The derivative of \re{stsm} with respect to $T$ gives
the heat capacity
 \be
 C_H=T\frac{dS}{dT}\approx
 \frac{2\pi^2(\widetilde{T})^3}{15\gamma^3}=\frac{2\pi^2{T}^3}{15c^3},
 \lab{cvsm}
\ee
which is common for interacting BEC systems since its measurement in superfluid helium \ci{helium}.  Note that for an ideal
Bose gas, i.e., for a system of  noninteracting particles,
the dispersion is not linear but quadratic, and $C_H\sim T^{3/2}$ \ci{ouriman}.

To find an expression for the Gr{\"u}neisen parameter, we use Eq.
\re{11.9} with the relations
\be
\Gamma_H=-\frac{1}{C_H}\left(\frac{dM}{dT}\right)_H=-\frac{g\mu_B}{C_H}\left(\frac{d\rho}{dT}\right)_H=
-\frac{g\mu_B}{2UC_H}\Delta_{T}^{\prime}.
\lab{grsm}
\ee
The expansion for $\Delta_{T}^{\prime}$ becomes
\be
\ba
\Delta_{T}^{\prime}=-\alpha_{1}\widetilde{T}-\alpha_{3}\widetilde{T}^3 +O(\widetilde{T}^5)
\lab{deltsm}
\ea
\ee
where
\bea
&&\alpha_1=\frac {8}{3}\frac{U}{UQ_{0}^{2}+4\,c},\\
&&\alpha_3=\frac{8U}{45\gamma^{2}}
\frac{3\pi^{2}UQ_{0}^{2}+12\pi^{2}c+10\,U{\gamma}^{2}}
{UQ_{0}^{2}+4\,c^{2}}.
 \lab{alpha13}
\eea

Inserting $C_H$ from \re{cvsm} we find
 \be
 \Gamma_H=
 \frac{15g\mu_{B}\alpha_{1}\gamma^{2}}{4\pi^{2}U}
 \frac{1}{\widetilde{T}^{2}}+
 \frac{15g\mu_B(2\gamma\alpha_3-\alpha_{1}^{2})}{8U\pi^{2}\gamma}+O(\widetilde{T}^2).
 \lab{ghsmex}
 \ee

 This is one of the central results of our paper. We will further simplify and discuss it later
 in the Discussion section (see Eqs.
 \re{ghcompact} to \re{ggr}).

 Using Eqs. \re{11.9}, \re{rhotiot} and \re{rho0}, the low-temperature
 expansions for the magnetizations become
 \be
 M=g\mu_B\rho(T)=M(0)-\dsfrac{g\mu_B \alpha_1}{4U\gamma m}\widetilde{T}^2+O(\widetilde{T}^4)
 \lab{Msm}
 \ee
 and
 \be
 \ba
 M^2_{\perp}=M^2_{\perp}(0)-\dsfrac{g^2\mu_{B}^{2}c(3\alpha_1+Um)}{24U\gamma^2}\widetilde{T}^2
  +O(\widetilde{T}^4).
  \ea
 \lab{Mstagsm}
 \ee

Both quantities  vary as  $-T^2$ in the low-temperature limit while
for a non-interacting Bose Einstein condensate, one has the $-T^{3/2}-$dependence.

Finally, we mention that the above relations for thermodynamic quantities
given in Eqs. \re{stsm}-\re{Mstagsm} are also valid in the HFP approximation,
but with slightly modified
\bea
&&\alpha_1\vert_{\rm HFP}=\dsfrac {8}{3}\dsfrac{U}{UQ_{0}^{2}+8c},\\
&&\alpha_3\vert_{\rm HFP}=\dsfrac{16U}{45\gamma^{2}}
\dsfrac{3\,{\pi }^{2}UQ_{0}^{2}+24\,{\pi }^{2}c+5\,U{\gamma}^{2}}
{UQ_{0}^{2}+8\,c^{2}}.
 \lab{alpha23}
\eea

\section{Properties near $T_c$}

The behavior of thermodynamic quantities in the temperature region $T\rightarrow
T_c \pm 0$ is crucial for the nature of a phase transition. According to
the Ehrenfest classification, a discontinuity in a second
derivative of $\Omega$ at $T_c$ with a continuous first derivative
indicates that the transition is of second order \ci{huangbook}. In
the present section we will
study $C_{H}^{(\pm)}\equiv C_H(T_c\pm 0)$, $\Gamma_{H}\equiv\Gamma_{H}(T_c\pm 0)$ and $S^{(\pm)}\equiv\,S^{(\pm)}(T_c\pm 0)$.

 \subsection{$T\rightarrow T_c  + 0$ region}
 Here ${\cE}_k=\omega_k=\veps_k $ and $\mu=2U\rho$. From Eqs. \re{cvbigtc} and
 \re{a1}-\re{a3} we have
 \be
 \ba
 C_{H}^{(+)}=-S_3+2US_1{{\rho^{\prime}}_T}
 \lab{cv1}
 \ea
 \ee
 where with $\beta_c=1/T_c$
 \be
 \ba
 S_3=-\beta_c^2\ds\sum_k\dsfrac
 {\veps_{k}^{2}e^{\beta_c\veps_k}
 }
 {(e^{\beta_c\veps_k}-1)^2
 },
 \lab{s3}\\
 S_1=-\beta_c\ds\sum_k\dsfrac
 {\veps_{k}e^{\beta_c\veps_k}
 }
 {(e^{\beta_c\veps_k}-1)^2
 },\\
 \lab{s1}
 {{\rho^{\prime}}_T}=\dsfrac
 {
 \beta_c S_1
 }
 {
 2S_2-1
 },\\
  S_2=-U\beta_c\ds\sum_k\dsfrac
 {e^{\beta_c\veps_k}
 }
 {(e^{\beta_c\veps_k}-1)^2
 }.
 \lab{S2}
 \ea
 \ee
 It can be easily shown that
 \be
 \lim_{T\rightarrow T_c  + 0}\rho^{\prime}_T=0
 \lab{rhosht}
 \ee
 since in this limit, $S_2$ in the denominator of Eq. \re{S2} at small momentum
 has an infrared divergence, since the integrand behaves as $k^{-2}$ in this limit. while the numerator is finite.
  Thus, Eq. \re{cv1} becomes
 \be
 C_{H}^{(+)}=-
  S_3=\beta_c^2\sum_k\dsfrac
 {\veps_{k}^{2}e^{\beta_c\veps_k}
 }
 {(e^{\beta_c\veps_k}-1)^2
 }.
 \lab{cvrt}
 \ee
 From Eq. \re{rhosht} we may immediately conclude that the Gr{\"u}neisen parameter at $T=T_c$
 vanishes,
 \be
 \Gamma_{H}^{(+)}=-
 \frac{1}{C_{H}^{(+)}}\lim_{T\rightarrow T_c  + 0}\left(\frac{dM}{dT}\right)_H=
 -\frac{g\mu_B}{C_{H}^{(+)}}\lim_{T\rightarrow T_c  + 0}\left(\frac{d\rho}{dT}\right)_H=0,
  \lab{gam0}
 \ee
 in agreement with the prediction of Garst \textit{et al.} \ci{garst}.
The entropy of Eq. \re{9.4} is with ${E}_k=\veps_k$
 \be
 S^{(+)}=-\sum_{k}
\ln\left[1-\exp(-\beta_c {\veps_{k}})\right]+\beta_c\sum_{k}
\frac{
\veps_{k}
}
{
(e^{\beta_c \veps_{k}}-1)
}.
 \lab{stpl}
 \ee

In the HFP approximation,   the equations \re{cv1} - \re{stpl}
 remain unchanged, since it coincides for $T\geq T_c$  with the HFB  approximation.

 \subsection{$T\rightarrow T_c  - 0$ region}

Here $\rho_0=0$, $\sigma=0$ and $\Delta=0$ and hence
${\cE}_k=E_k=\veps_k$ again, i.e.,
the dispersion is the same on both sides of the critical
temperature. For this reason the entropy is continuous at $T=T_c$,
$S^{(-)}=S^{(+)}$.
The heat capacity and the Gr{\"u}neisen parameter are
 \be
  C_{H}^{(-)}=
  \beta_c^2\sum_k\dsfrac
 {
 \veps_{k}(\veps_{k}-T_c\Delta^{\prime}_T)e^{\beta_c\veps_k}
 }
 {
 (e^{\beta_c\veps_k}-1)^2
 },
 \lab{cvminus}
  \ee
\be
  \Gamma_{H}^{(-)}=-
 \frac{g\mu_B\Delta^{\prime}_T}{2UC_{H}^{(-)}},
 \lab{grunminus}
 \ee
 where we used the relation $\rho^{\prime}_T=\Delta^{\prime}_T/2U$. The  $\Delta^{\prime}_T$ defined in \re{bdelta}
 for the HFB approximation may be rewritten as
 \be
 \ba
\Delta^{\prime}_T\vert_{\rm HFB}=\dsfrac{\beta_c U S_4}
{2(2S_5+1)},\\
S_5=-U\beta_c\sum_{k}\dsfrac
{
T_c+\veps_ke^{\beta_c\veps_k}-T_ce^{\beta_c\veps_k}
}
{
\veps_k(e^{\beta_c\veps_k}-1)^2
},
\\
S_4=-4\beta_c\sum_{k}\dsfrac
{
\veps_ke^{\beta_c\veps_k}
}
{
(e^{\beta_c\veps_k}-1)^2
}.
 \lab{deltapr}
 \ea
 \ee
In the HFP approach,
\be
\Delta^{\prime}_T\vert_{\rm HFP}=\dsfrac{2\beta_c U S_1}
{2S_{2}+1},
 \lab{deltaprHFP}
 \ee
where $S_1$ and $S_2$ are the same as in
Eqs. \re{s1}.

\subsubsection{HFB approximation}
From Eqs. \re{cvrt}, \re{gam0} and \re{cvminus} we can express
the discontinuities in $C_H$ and $\Gamma_H$ as
\be
\ba
\Delta C_H=C_{H}^{(-)}-C_{H}^{(+)}=
-\beta_c\sum_{k}\dsfrac
{
\veps_k\Delta^{\prime}_T e^{\beta_c\veps_k}
}
{
(e^{\beta_c\veps_k}-1)^2
}>0,\\
\lab{jumpch}
\ea
\ee
\be
\ba
\Delta \Gamma_H=\Gamma_{H}^{(-)}-\Gamma_{H}^{(+)}=
-\dsfrac
{
g\mu_B\Delta^{\prime}_T
}
{
2UC_{H}^{(-)}
}>0,
\lab{jmpgh}
\ea
\ee
where $\Delta^{\prime}_T$ is given in \re{deltapr} and $C_H$ in \re{cvminus}.

From Eqs. \re{jumpch} and \re{jmpgh} it is clear that not only $C_H$ but
also the Gr{\"u}neisen parameter
has a finite jump near the critical temperature, and therefore the transition is of
second order according to the Ehrenfest classification.

It is easy to show that our results satisfy self-consistently the Ehrenfest relation
\be
\Delta C_H=-T_c\left(\frac{dH_c}{dT}\right)_{T=T_c}\Delta
\left(\frac{dM}{dT}\right)_{T=T_c}.
\lab{erf}
\ee
Using equations \re{10.4}, \re{11.9}, \re{rhosht} and \re{jumpch}
leads to a modified Ehrenfest relation for the discontinuity in the Gr{\"u}neisen
parameter in triplon systems,
\be
\Delta \Gamma_H=\dsfrac{\Delta C_H}{T_c C_{H}^{(-)} (dH_c/dT)      },
\lab{erandr}
\ee
which can be easily derived  from Eqs.  \re{10.4}, \re{cvminus},  \re{jumpch} and
\re{jmpgh}.

\subsubsection{HFP approximation}

Here $\Delta^{\prime}_T$ is given by Eq. \re{deltaprHFP}, where $S_1$ is finite
but $S_2$ diverges as it has been shown in the previous section. Thus
$\Delta^{\prime}_T\vert_{\rm HFP}=0$, and hence
\be
\rho^{\prime}_T\vert_{\rm HFP}=(\frac{\Delta+\mu}{2U})^{\prime}=0.
\lab{rhosthfp}
\ee
Therefore, we may conclude from Eqs. \re{cvrt}, \re{cvminus} and \re{rhosthfp}
\be
\ba
 C_{H}^{(-)}\vert_{\rm HFP}= C_{H}^{(+)}\vert_{\rm HFP}\quad \quad
  \Gamma_{H}^{(-)}\vert_{\rm HFP}= \Gamma_{H}^{(+)}\vert_{\rm HFP}.
\lab{chgrunnojump}
\ea
\ee
 In other words, there is \textit{no discontinuity}  in the HFP approximation, neither in the heat capacity
 nor in the Gr{\"u}neisen parameter, which is  in sharp contrast to
 the HFB approximation used here, and to experimental heat-capacity
 measurements.

 \section{Discussion}

 \subsection{Sign change of $\Gamma_H$ at $T_c$}

 In the previous section, we have shown that $\Gamma_H = 0$ at the
 critical temperature $T_c$. It is also easy to show that $\Gamma_H(T)$ must
 change its sign there. Using Eqs. \re{gam0} and \re{a3} we have with
 $d\rho/dT > 0$ a $\Gamma_H(T) < 0$ for $T >T_c$. Approaching the
 critical temperature from below where $d\rho/dT < 0$ (see Eq.
 \re{grunminus}), $\Gamma_H(T) > 0$ for $T <T_c$.

\subsection{Divergence of $\Gamma_H$ near the transition}
 Rewriting Eq. \re{ghsmex} around the QCP in the limit
  $r=(H-H_c)/H_c \rightarrow 0$ in a compact form (see Appendix C), we obtain
  \be
  \Gamma_H\approx \frac{G_t (H-H_c)}{T^{2}}+\frac{G_r}{H-H_c},
  \lab{ghcompact}
  \ee
  with
  \be
  G_t=\frac{10g^2\mu_{B}^{2}}{UmQ_{0}^{2}\pi^2}=\frac{5g^2\mu_{B}^{2}}{\pi^2}G_r
  \lab{ggt}
  \ee
  and
   \be
   G_r=\frac{2}{UmQ_{0}^{2}},
   \lab{ggr}
   \ee
  where the next higher-order terms are $O((\widetilde{T})^2)$ and $O(r)$,
  respectively. Here we used the relation
  \be
  U=\frac{4\pi a_s}{m},
  \lab{as}
  \ee
   where $a_s$ is the s- wave  scattering length.
The first term in Eq. \re{ghcompact} dominates $\Gamma_H(T,H)$ in a fixed
  magnetic field $H > H_c$ for temperatures
  $T\ll\eta(H-H_c)$, with $\eta = \sqrt{5}g\mu_{B}/\pi$, while the second term
  dominates in the opposite limit when $H$
  approaches the QCP $H_c$ from above at a fixed low temperature $T$.

  The fact that $\Gamma_H$ diverges as $\Gamma_H=1/T^2$ at low
  enough temperatures is one of our main results.
  Remarkably, the classification of a number of magnetic systems ranging from heavy-fermion
  compounds to frustrated magnets done by Gegenwart \textit{et al.} \cite{gegenw2} reveals that the majority
  of considered systems shows indeed a similar behavior, with some exceptions like \ci{dtn2012}, however.

  The phase boundary between the condensed and the
  uncondensed states in spin gapped quantum magnets, respectively, can be expressed by a power law
  of the form $T_c\propto (H-H_c)^\phi$. As experimental data on insulating spin systems often show
  $\phi\approx 0.5$ (see Table \ref{tab1} in the next section), and our results are in line with experimental
  observations.

The behavior $\Gamma_{H}\simeq G_{r}(H-H_{c})^{-1},$ being well
established \cite{gegenw2,garst} in the QCP systems, is obviously also realized in the
systems discussed in the present work (see Eq. \re{ghcompact}).
We note, however, that this relation cannot be directly applied to the continuous systems such  as
atomic gases where $Q_0\rightarrow \infty$. In this case, renormalization procedures may lead to
different dependences.

\subsection{Universality of $G_r$}

We now discuss whether the value of the dimensionless
parameter $G_r$ is universal in spin-gapped triplon systems or not.
We make use of our result \re{ggr}, and state that $G_r$ only depends on the
product of the material parameters $U$ and $m$. We will now show that
within our assumptions, $U$ and $m$ are not really independent of each other.
The full width of the model dispersion relation \re{bare} is $D = 3J_0 = 3/m$.
The lower and the upper bounds of the gapped lowest magnon band
with bandwidth $D$ determine, in a crude approximation, the width of the magnetic phase
(i.e., the values of $H_{c1}$ and $H_{c2}$,
respectively) by the Zeeman shift of the corresponding lowest and
highest lying triplet states, respectively \cite{giam}.
Although our approach is only valid in the dilute limit near $H_{c1}$,
we can formally extrapolate it to the fully polarized state with $M = g\mu_B$
at $H_{c2}$, and then $\mu(H_{c2})-\mu(H_{c1})=\mu(H_{c2})\approx D=3J_0$.
The state at $H_{c2}$ corresponds with Eq. \re{11.9} to $\rho = \rho_0 = 1$ at $T = 0$.
With Eq. \re{11.3} we find $\mu(H_{c2}) \approx U \approx 3J_0$, and
therefore $Um \approx 3$.

  In real systems, however, the dispersion relation will deviate from Eq. \re{bare},
  and the bandwidth $D$ then differs from the value $3/m$ containing the
  low-energy effective mass $m$. Moreover, the
  magnon bands are not rigidly shifted by the Zeeman effect
  in magnetic fields between $H_{c1}$ and $H_{c2}$ \cite{matsumoto},
  so that $U \approx D \approx 3J_0$ can hold only by up to a factor of
  unity. In this sense, relations \re{ggt} and \re{ggr} are not
  strictly universal, but depend on the details of the magnon spectrum.
  In Table \ref{tab1} to be presented below, we have indeed a ratio $U/J_0=Um$
  between 3.2 - 6.3, deviating from the value of 3.

 \subsection{Numerical results for real systems}

In the previous sections, we have given general expressions for $\Gamma_H$, $S$, $C_H$, $M$ and $M^2_{\perp}$, and
elaborated the limiting cases $T\rightarrow 0$ and $T\rightarrow T_c$.
We can use these results to numerically evaluate these
quantities over the full range of temperatures. In the following we will restrict
ourselves to $\Gamma_H$ and $S$. To do this, we have to assume a set
of realistic material parameters $g$, $H_{c}$, $U$ and $J_0$ which we
take from experimental data for Ba$_{3}$Cr$_{2}$O$_{8}$,
Sr$_{3}$Cr$_{2}$O$_{8}$ and TlCuCl$_{3}$
\cite{misgu,nikuni,kofuprl,aczelSr,wangprl} (see Table \ref{tab1}).

\begin{figure}[h!]
\begin{minipage}[H]{0.49\linewidth}
\center{\includegraphics[width=1.25\linewidth]{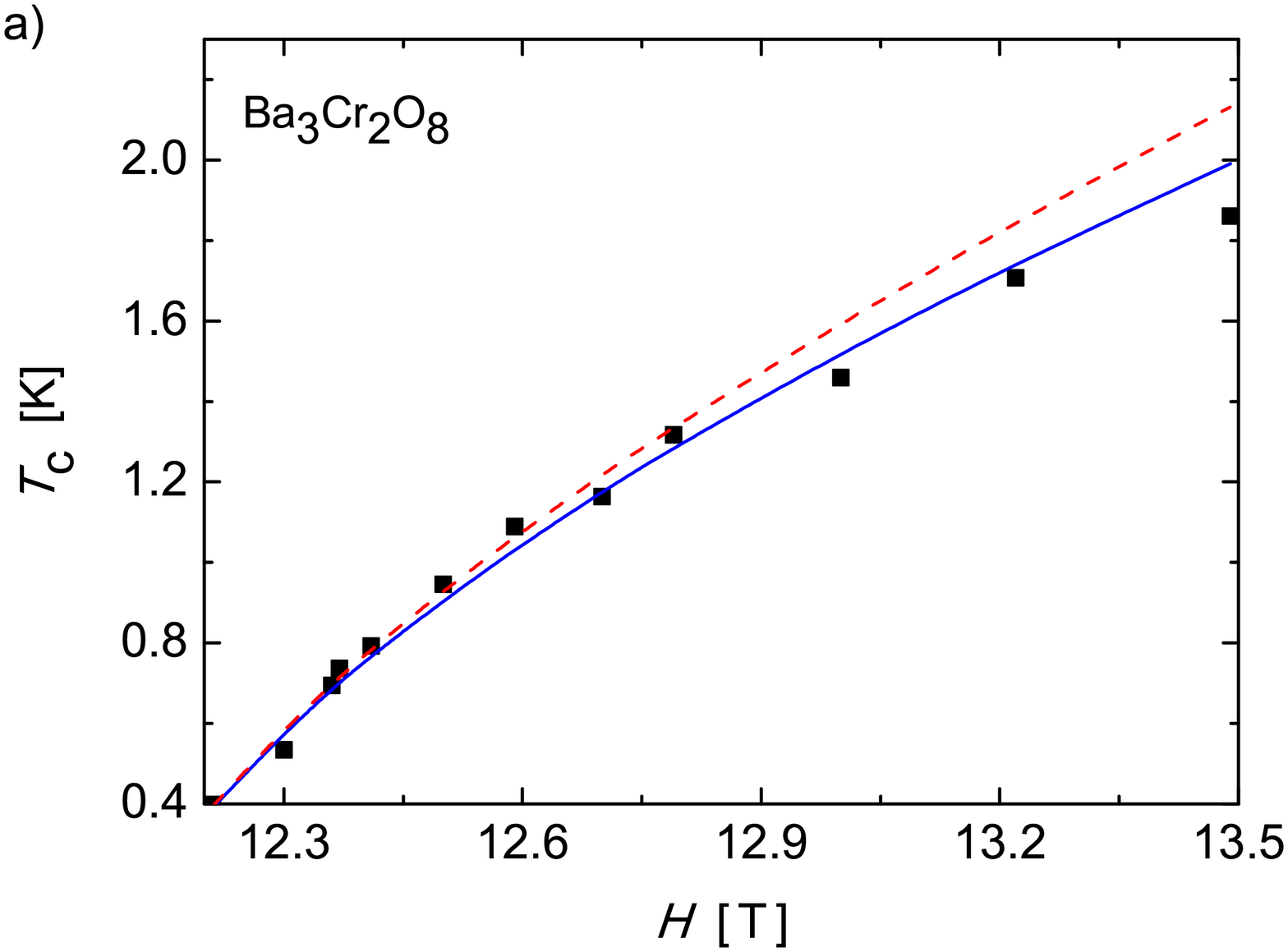} \\ a)}
\end{minipage}
\hfill
\begin{minipage}[H]{0.49\linewidth}
\center{\includegraphics[width=1.25\linewidth]{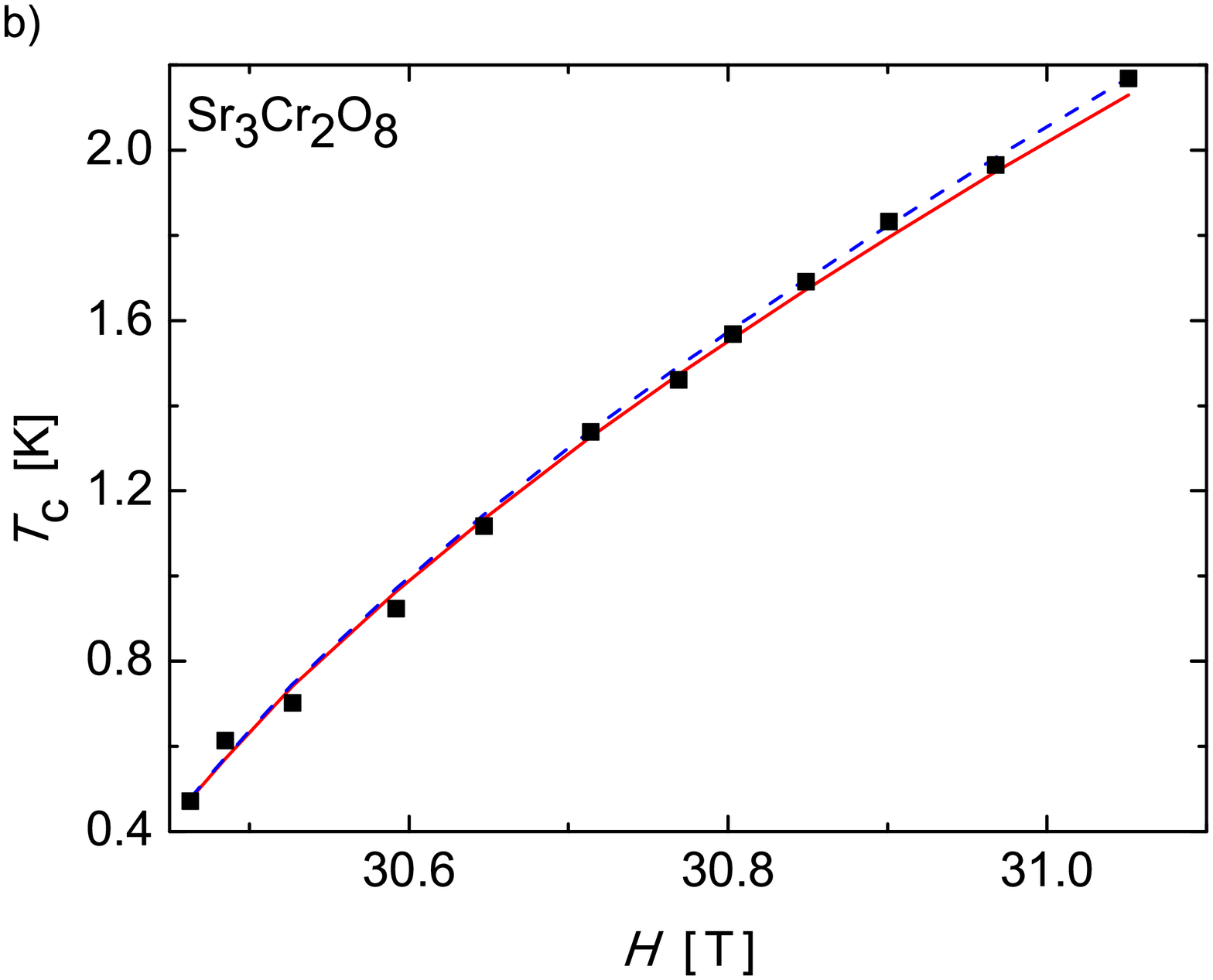} \\ b)}
\end{minipage}
\caption{The dependence of $T_{c}$ on
 the external magnetic field $H$ for (a) Ba$_{3}$Cr$_{2}$O$_{8}$
 and  (b) Sr$_{3}$Cr$_{2}$O$_{8}$ (solid lines from Eq. \re{10.4}).
 The dashed lines correspond to the $\phi =2/3$ law.
 The
 experimental data are taken from (a) \cite {kofuprl}
  and (b)
\cite{aczelSr,wangprl}.
}
\label{figTcH}
\end{figure}

To begin with, we show in Fig. \ref{figTcH} the phase diagrams $T_{c}(H)$ as calculated from Eq.
(\ref{10.4}) for Ba$_{3}$Cr$_{2}$O$_{8}$ and Sr$_{3}$Cr$_{2}$O$_{8}$,
together with experimental data taken from Refs. \cite{kofuprl,aczelSr,wangprl}.
For our calculation, we fixed $g$, $H_{c}$, and $U$, and fitted $J_0$ according to Eq. \re{10.4}.

In Table \ref{tab1}, we compare the exponent $\phi$ as obtained from
a power-law fit according to $T_c\propto (H-H_c)^\phi$ to our numerically obtained data,
with corresponding fits to the experimental data in the same temperature range ($\phi_{exp}$)
and to a range of values for TlCuCl$_{3}$ from the literature \cite
{nikuni,tanaka}. These exponents are in fair agreement with our expectation $\phi=\nu
z = 1/2$.

\begin{table}[h!]
\begin{tabular}{|p{4cm}|c|c|c|c|c|c|c|c|c|}
 \hline
    {}                    &   $g$   & $H_{c}(T)$
&$J_{0}=1/m(K)$&  $U(K)$ &
$\Delta_{\rm st}(K)$&$G_r$&$\phi$&$\phi_{\rm exp}$&$a_s/\bar{a}$ \\
 \hline
 Ba$_{3}$Cr$_{2}$O$_{8}$  &  1.95 &  12.10     & 5.045
&  20     &  15.85&0.84&0.5& 0.49&0.315 \\
 \hline
 Sr$_{3}$Cr$_{2}$O$_{8}$  &  1.95 &  30.40       &   15.86     & 51.2      &   39.8&0.9&0.65& 0.65 &0.257\\
 \hline
  TlCuCl$_{3}$  &  2.06 &  5.1       &   50   & 315
&   7.1&0.72 &0.62& 0.45-0.71 &0.5             \\
 \hline
   \end{tabular}
   \caption {Material parameters used for our numerical calculations.
   From the input parameters $g$, $H_{c}$ and $U$ we derived $J_{0}$
   from fitting the experimental phase boundary $T_{c}(H)$ to Eq. (\ref{10.4}).
   $\Delta_{\rm st}$ corresponds
   to the energy scale of $H_{c}$ in Kelvin, while $G_r$ and $a_s/\bar{a}$
   come from Eqs. (\ref{ggr}) and (\ref{as}). The exponents $\phi$ and $\phi_{exp}$ are results
   from fitting our numerically generated and experimental $T_{c}(H)$
   data, respectively, to a power law.}
   \lab{tab1}
\end{table}

\begin{figure}
\begin{minipage}[H]{0.49\linewidth}
\center{\includegraphics[width=1.1\linewidth]{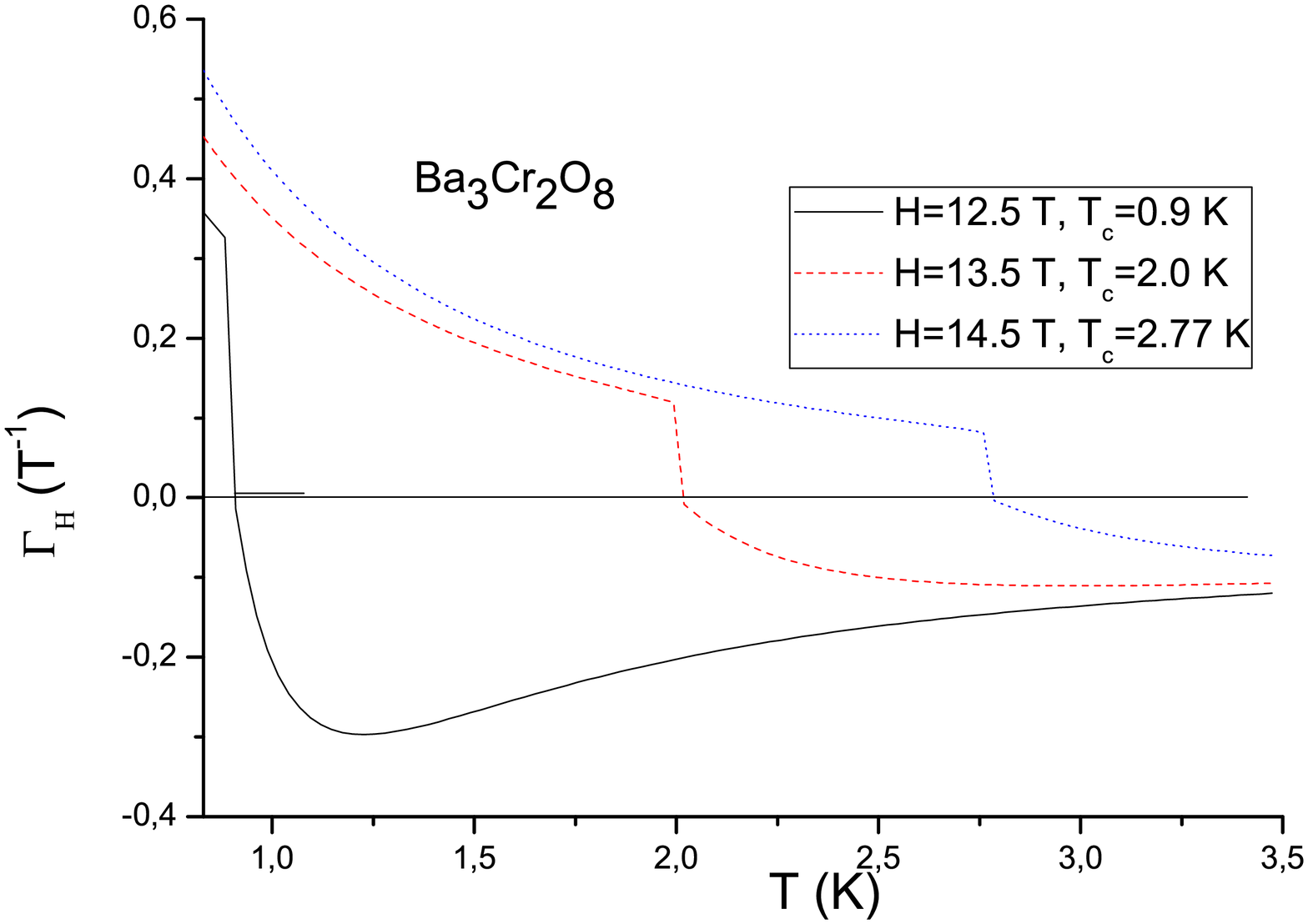} \\ a)}
\end{minipage}
\hfill
\begin{minipage}[H]{0.49\linewidth}
\center{\includegraphics[width=1.1\linewidth]{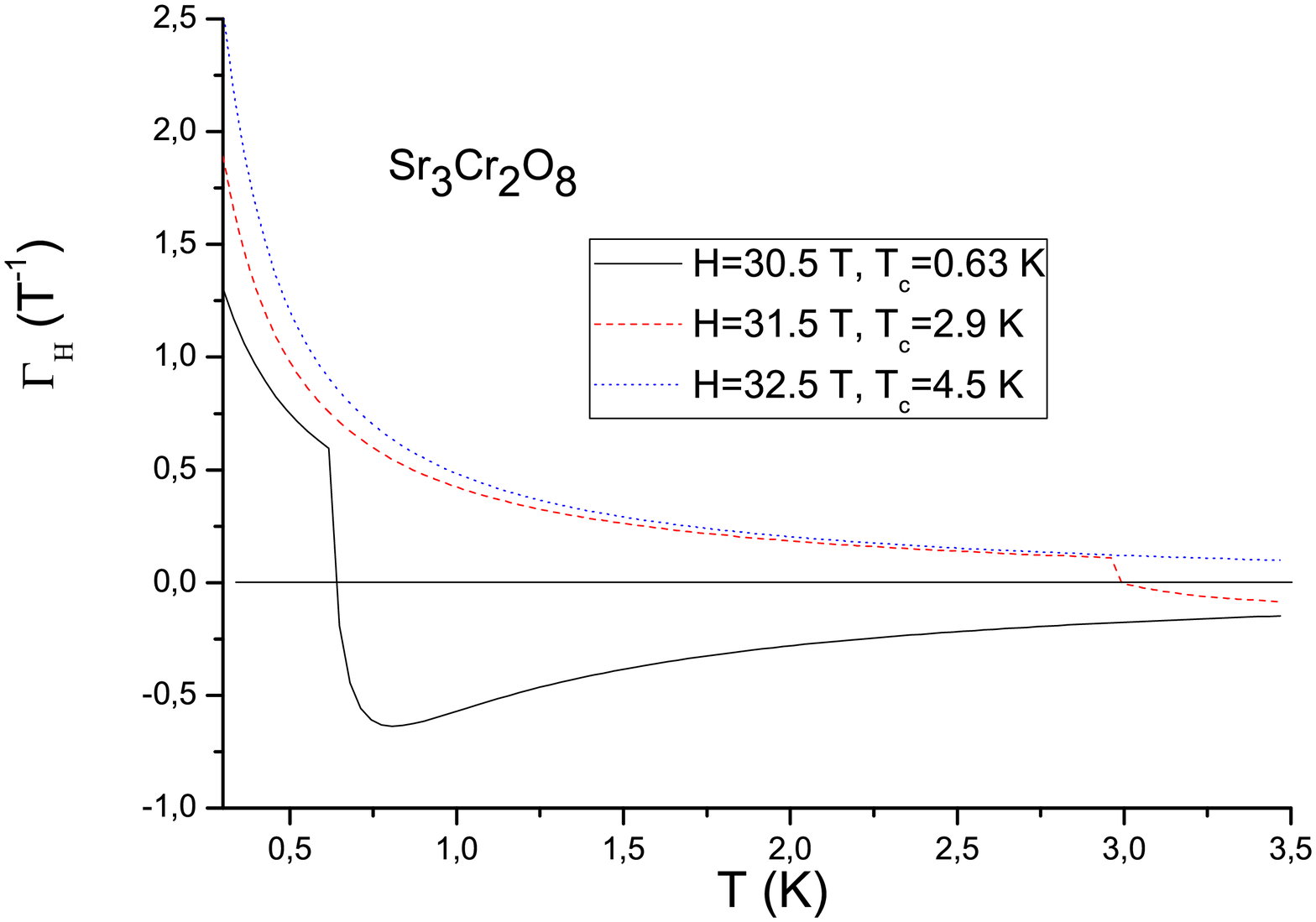} \\ b)}
\end{minipage}
\caption{
The dependence of the Gr{\"u}neisen parameter on
 temperature for (a) Ba$_{3}$Cr$_{2}$O$_{8}$
 and  (b) Sr$_{3}$Cr$_{2}$O$_{8}$ in different magnetic fields. At
 the respective $T_c$, $\Gamma_H(T)$ shows a discontinuity and changes its sign.
}
\label{GHT}
\end{figure}

\begin{figure}
\begin{minipage}[H]{0.49\linewidth}
\center{\includegraphics[width=1.1\linewidth]{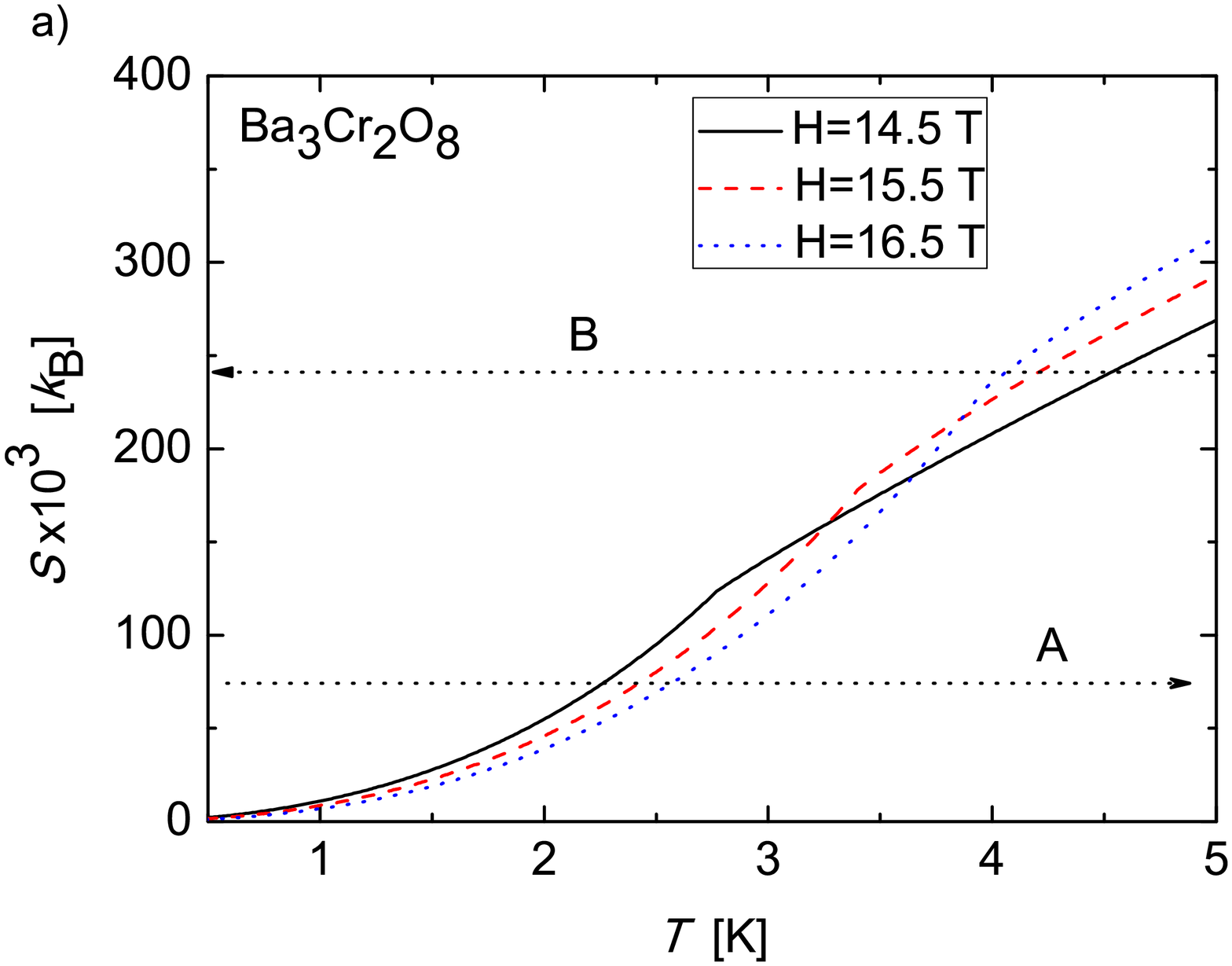} \\ a)}
\end{minipage}
\hfill
\begin{minipage}[H]{0.49\linewidth}
\center{\includegraphics[width=1.1\linewidth]{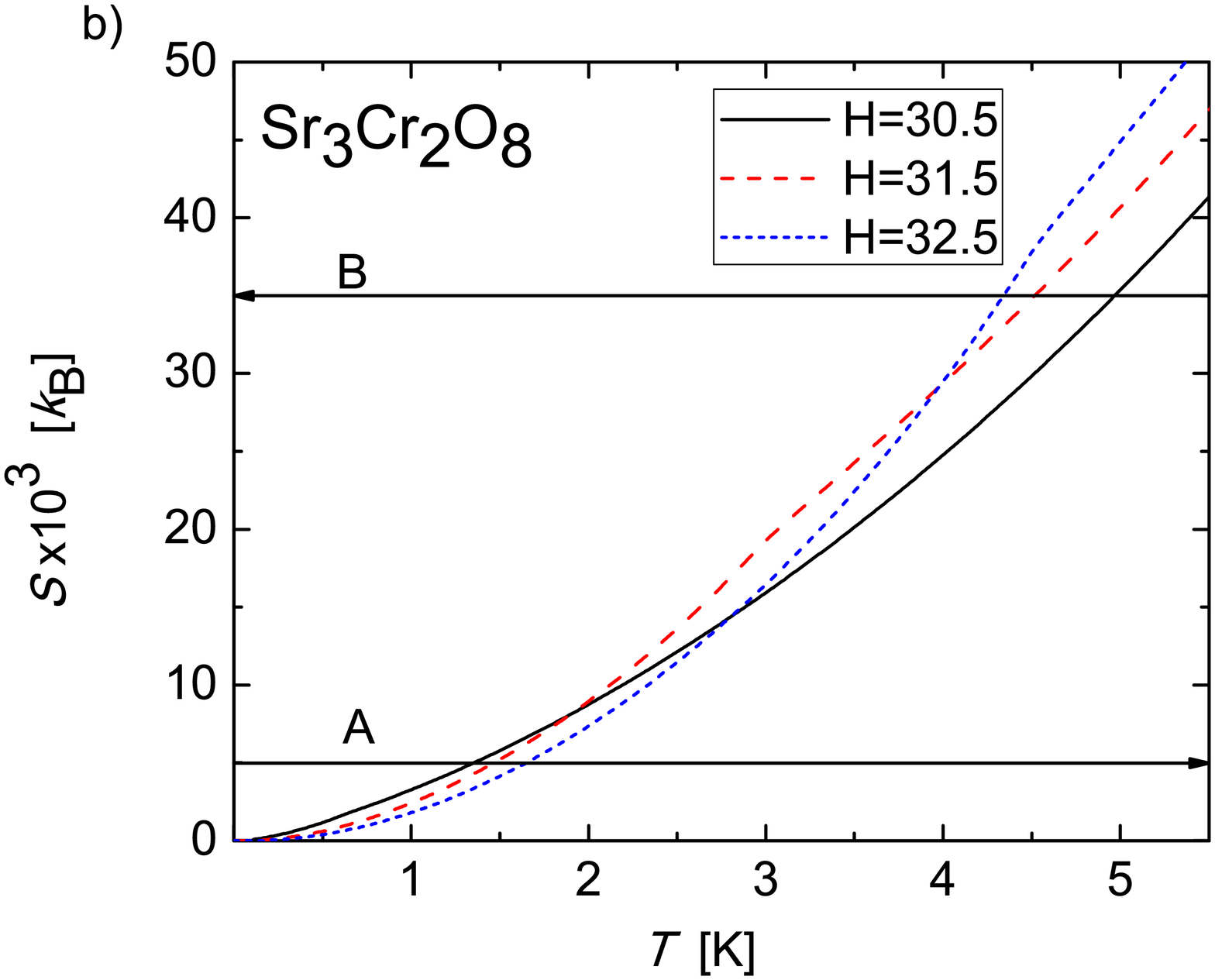} \\ b)}
\end{minipage}
\caption
{The entropy $S$ $vs.$ temperature $T$ for Ba$_{3}$Cr$_{2}$O$_{8}$
 and Sr$_{3}$Cr$_{2}$O$_{8}$ for different values of the magnetic field $H$.
 As expected, $S(T)$ changes its slope at $T_c$.
 }
\label{STemp}
\end{figure}

\begin{figure}
\begin{minipage}[H]{0.49\linewidth}
\center{\includegraphics[width=1.1\linewidth]{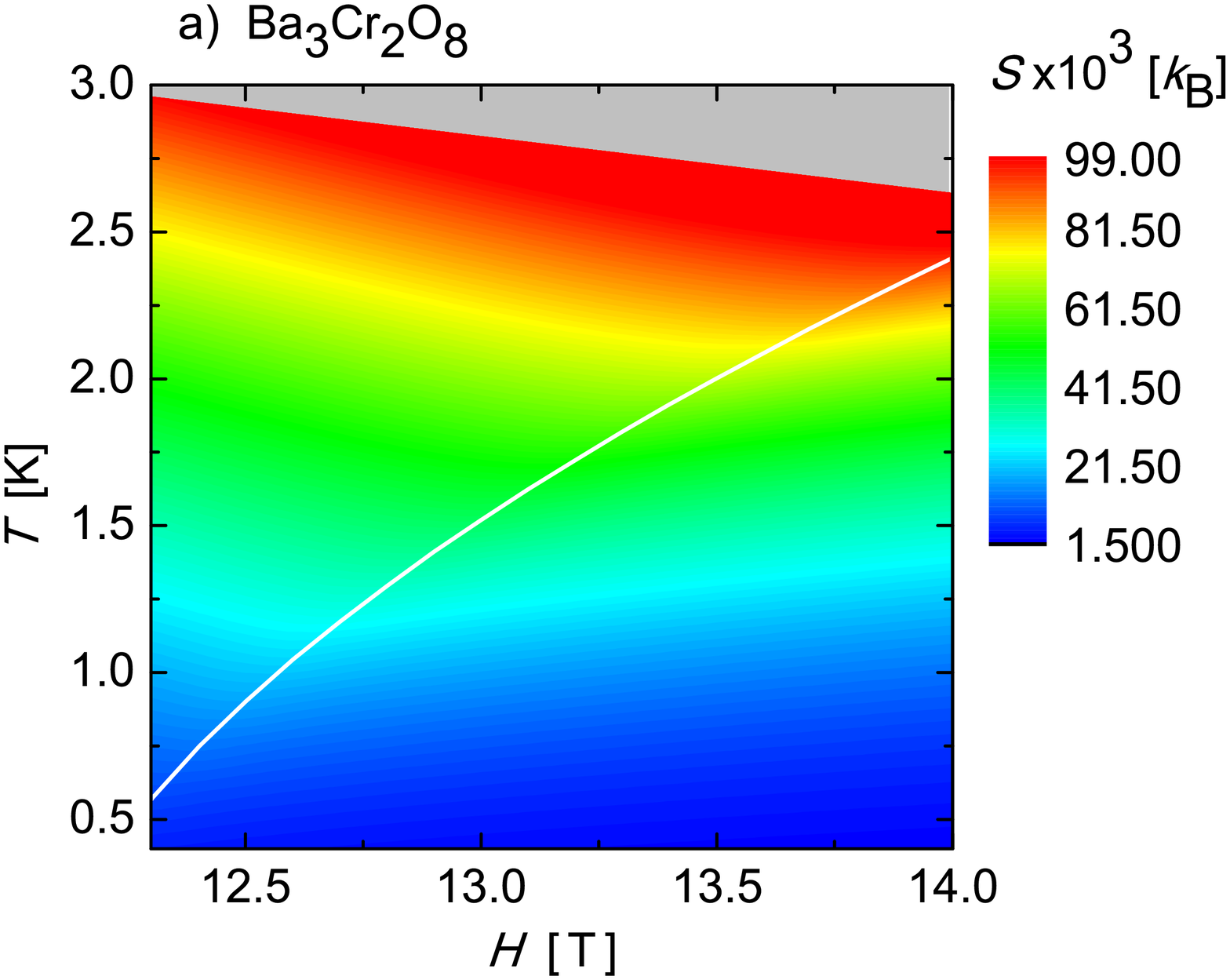} \\ a)}
\end{minipage}
\hfill
\begin{minipage}[H]{0.49\linewidth}
\center{\includegraphics[width=1.1\linewidth]{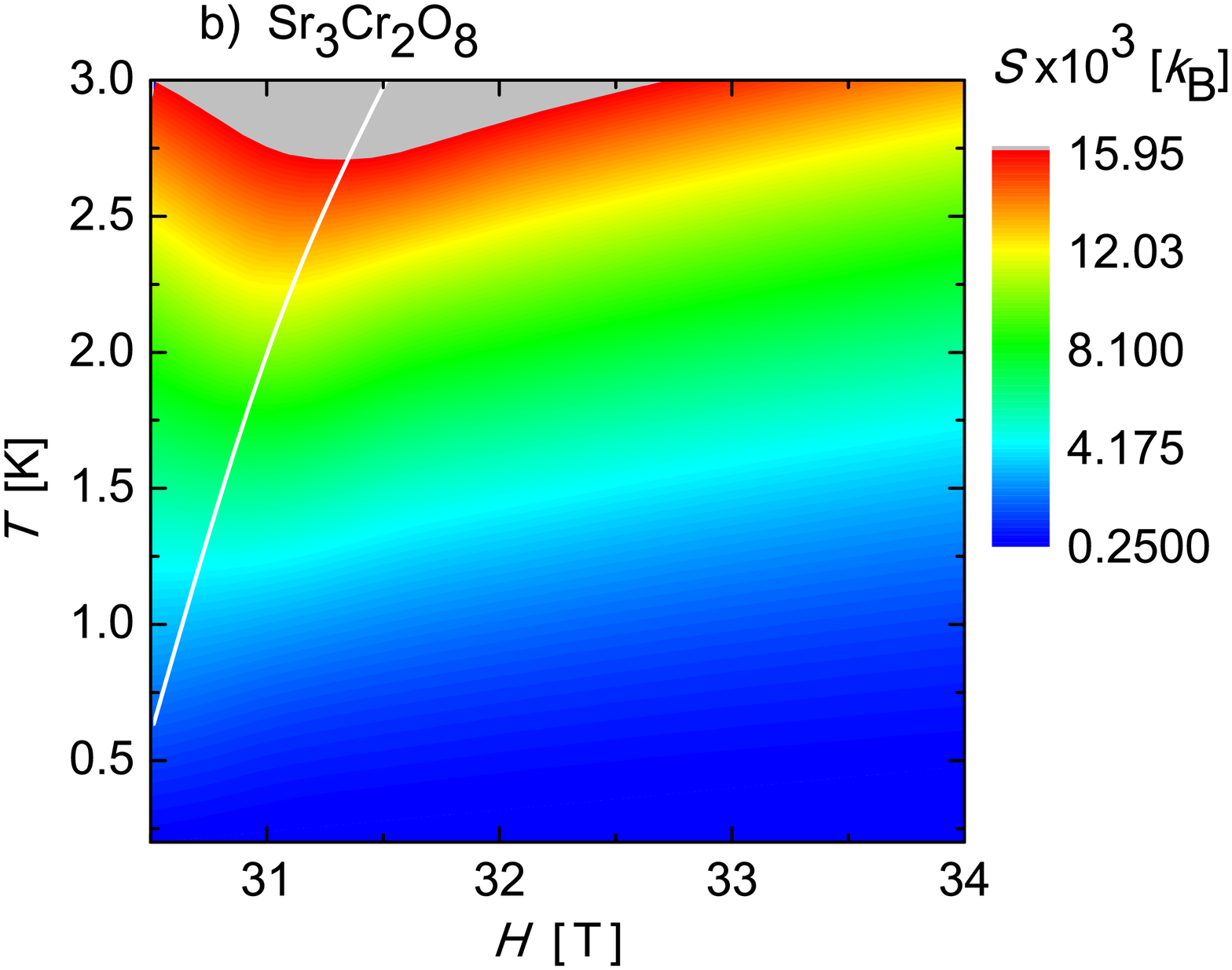} \\ b)}
\end{minipage}
\caption
{Isoentropic lines for the magnetic system of (a) Ba$_3$Cr$_2$O$_8$ and (b)
Sr$_3$Cr$_2$O$_8$ in the $(H,T)$ plane. Each color corresponds to a
constant entropy value. The white lines show the phase
boundaries separating the condensed (right side) from the uncondensed
phases (left side), respectively.
 }
\label{isentrop}
\end{figure}

Corresponding calculations for $\Gamma_H(T)$ using Eqs. \re{grtbig} and
\re{grsmtc} are shown in Fig. \ref{GHT} and for $S(T)$ in Fig.
\ref{STemp},
while in Fig. \ref{isentrop}, we display a series of isoentropic lines with
$S={\rm const}$ for Ba$_{3}$Cr$_{2}$O$_{8}$ and Sr$_{3}$Cr$_{2}$O$_{8}$.

The phase transition is clearly visible in all of these figures. The Gr{\"u}neisen parameter
$\Gamma_H(T)$ shows a discontinuity according to Eq. \re{erandr} and changes
its sign at $T_c(H)$, while the entropy $S(T)$ exhibits a change in
its slope, thereby reflecting a discontinuity in the heat capacity $C_H$
according to Eq. \re{jumpch}.

The isoentropic lines shown in Fig. \ref{isentrop} have a minimum at $H_c(T)$, which can be easily
understood by recalling that $\Gamma_{H}=T^{-1}(dT/dH)$ vanishes at
the phase transition. In a
perfectly adiabatic experiment, the temperature would ideally follow these lines upon a change
of the external magnetic field, reaching its lowest temperature at $H_c$.
The diagram shown in Fig. \ref{isentrop} for Sr$_3$Cr$_2$O$_8$
compares favorably with that measured by Aczel \textit{et al.}
\cite{aczelSr}. We note that in most conventional magnetocaloric experiments, a sample
is subject to a controlled heat link, so that the corresponding $T(H)$
curves become time-dependent \cite{schill,kohamamce,aczelSr,zapf} and
change their shape in comparison with those displayed in Fig. \ref{isentrop}.

\section{Measurability}

While the discontinuities in $C_H$ and $\Gamma_H$ (Eqs. \re{jumpch} and \re{jmpgh}) and the sign change in
$\Gamma_H$ at $T_c$ can, in principle, be directly measured in a dedicated experiment, an
examination of the temperature dependence of these quantities in the
low-temperature limit may face the problem that the
heat capacity $C_H$ of the magnetic subsystem exhibits the same temperature
dependence as that of the crystal lattice, i.e., $C_H\sim T^{3}$, and
the magnetic contribution has then to be extracted
from the total signal. This is possible, e.g., by performing a series
of measurements in different magnetic fields (and in $H=0$
in the case of  $C_H$) provided that the lattice heat capacity
$C_{\rm lat}$ does not entirely dominate $C_H$. For $N_{\rm at}$
atoms in the crystal lattice,
we have
in the low-temperature limit the Debye result

\be
 C_{\rm lat}\approx N_{\rm at}\frac{12\pi^4{T}^3}{5\Theta_{D}^3},
\lab{latt}
\ee
with the Debye temperature of the lattice $\Theta_D$.
The number of dimers $N_{\rm dim} < N_{\rm at}$ enters as a prefactor in Eq.
\re{cvsm} to express the heat capacity of the whole system, and the ratio of
the two contributions then becomes

\be
\frac{C_H}{C_{\rm lat}} = \frac{1}{18\pi^2}\frac{N_{\rm dim}}{N_{\rm at}}\frac{\Theta_{D}^3}{c^3},
\lab{ratio}
\ee
where $c$ from $E_k=ck$ is expressed in Kelvin. This ratio seems
to be unfavorably small. However, by performing an experiment close
enough to the QCP one can force $c \ll \Theta_D$, and the two
contributions may become separable.\footnote{With a $J_0$ = 15 K and
$g \approx 2$ as for Sr$_3$Cr$_2$O$_8$, we estimate $c \approx  4.5$ K   for $H - H_c = 1$ T, so that
with $N_{\rm dim}/N_{\rm at} = 1/13$ and $\Theta_D \approx 120$ K
 \cite{wangprb}, $C_H/C_{\rm lat} \approx 8$} The qualitative field dependence $\Gamma_{H} \sim (H-H_{c})^{-1}$ from
Eq. \re{11.9} and the magnetizations \re{ghcompact}, \re{Msm}
and \re{Mstagsm} remain unaffected by these arguments and should be readily accessible
in a corresponding experiment, while the absolute value of the
measured $\Gamma_H$ has to be corrected for the contribution of
$C_{\rm lat}$ to Eq. \re{Grun}.

 \section{Conclusion}

 We have performed a variational Gaussian-approximation analysis of gapped dimerized
 quantum magnets showing a Bose-Einstein condensation of magnetic quasiparticles (triplons).
 We calculated the free energy $\Omega$ and the associated entropy $S$, the heat capacity $C_H$,
 the magnetization $M$ and the Gr{\"u}neisen parameter $\Gamma_H$, and derived explicit expressions for
 these quantities in the limits $T\rightarrow T_c$ and $T\rightarrow 0$, respectively. Near the critical temperature,
 both the heat capacity and the Gr{\"u}neisen parameter show a discontinuity, while $\Gamma_H$
 also changes its sign. Such a behavior is expected for systems with a magnetically
 controlled quantum critical point \cite{garst}. In the low-temperature limit near this QCP,
 we find that $C_H \sim T^3$, which is universal for Bose condensed interacting systems.
 The Gr{\"u}neisen parameter diverges there as $\Gamma_H \sim T^{-2}$ as a function of temperature.
 To the best of our knowledge this is a new result,
 and we have confirmed that a corresponding experiment to verify this conjecture
 should be feasible. Approaching the transition field
 as $H\rightarrow H_c$ we find $\Gamma_H \sim (H-H_c)^{-1}$, which is common
 to a variety of magnetic systems \cite{gegenw2}. We have also shown that the Gr{\"u}neisen parameter and heat capacity are continious near critical temperature 
 in the HFP approximation which is in contrast to experemintal heat capacity 
 measurements.

\section*{Acknowledgments}

We are indebted to Evgeny Sherman and Philipp Gegenwart  for useful discussions.  This work is partially supported by the Swiss National Foundation SCOPES project
IZ74Z0\_160527. 

\newpage

\section*{Appendix A  }
\def\theequation{A.\arabic{equation}}
\setcounter{equation}{0}

Here we derive the free energy given in \re{8.1} by using a variational perturbative theory, which is similar to HFB approach in Hamiltonian formalism \cite{andersen}. 
This perturbative scheme includes the following
  steps:

1)  We parameterize the quantum filed $\psi$ in terms of a time-
  independent condensate ${\rho_0}$ and a quantum fluctuation
  field $\widetilde{\psi}$ as
  \be
  \ba
  \psi=\sqrt{\rho_0}+\widetilde{\psi}
  \ea
  \ee
  which defines the number of uncondensed particles as
    \be
   \rho_{1}=\int d^{3}r\langle\widetilde{\psi}^{\dag}\widetilde{\psi}
    \rangle,
    \label{6.2}
    \ee
 where the expectation value of an operator $\langle \hat{O}(\widetilde{\psi}^{\dag}, \widetilde{\psi})\rangle$
 is defined as
\be
 \langle\hat{O}\rangle=\frac{1}{\cal Z}
 \int{\cal D}\widetilde{\psi}^{\dag}{\cal D}
  \widetilde{\psi} \hat{O}(\widetilde{\psi}^{\dag}, \widetilde{\psi})e^{-{\cal A}
  [\widetilde{\psi}^{\dag},\widetilde{\psi}]}.
   \label{6.3}
   \ee
   Then the  total number of particles is given by
   \be
    \rho=\rho_{1}+\rho_{0}.
    \label{6.4}
    \ee

2) We replace $U$ in (\ref{2.1}) as $U\to\delta U$ and add to
(\ref{2.1}) following term:
\be
 S_{\Sigma}=(1-\delta)\int d\tau d^{3}r \left[
  \Sigma_{\rm n}\widetilde{\psi}^{\dag}\widetilde{\psi}+
   \frac{1}{2}\Sigma_{\rm an}(\widetilde{\psi}^{\dag}\widetilde{\psi}^{\dag}
   +\widetilde{\psi}\widetilde{\psi})\right],
      \ee
   where the variational  parameters $\Sigma_{\rm n}$ and $\Sigma_{\rm an}$
    may be interpreted as the normal and the anomalous self energies,
    respectively.

3) Now the perturbation scheme may be considered as an expansion
  in powers of $\delta$ by using the propagators
 \be
 G_{ab}(\tau,r;\tau^\prime,r^\prime)=\frac{1}{\beta}\sum_{n,k}
 e^{i\omega_{\rm n}(\tau-\tau^\prime)+i\bfk(r-r^\prime)} G_{ab}(\omega_{\rm n},
 \bfk)
  \label{7.1}
  \ee
 $(a,b=1,2)$, where $\omega_{\rm n}=2\pi nT$ is the $n$-th Matsubara
  frequency, $\displaystyle{\sum_{n,k}=\sum_{n=-\infty}^{\infty}
 \int d^{3}k/(2\pi)^3}$, and
  \begin{eqnarray}
   G_{ab}(\omega,\bfk)=\frac{1}{\omega_{\rm n}^{2}+{E}_k^{2}}
    \left(
\begin{array}{lr}
 {\varepsilon}_{k}+X_{2} & \omega_{k}\\
 -\omega_{k} &{\varepsilon}_{k}+X_{1}
\end{array}\right).\label{7.2}
  \end{eqnarray}

 In (\ref{7.2}) ${E}_k$ corresponds to the dispersion of quasiparticles
 \be
 {E}_k=\sqrt{\varepsilon_{k}+X_{1}}\sqrt{
 \varepsilon_{k}+X_{2}},
 \label{7.21}
 \ee
 where $X_{1}$ and $X_{2}$, given by
 \be
  \ba
  \displaystyle{
  X_{1}=\Sigma_{\rm n}+\Sigma_{\rm an}-\mu},\nonumber\\
  \displaystyle{
  X_{2}=\Sigma_{\rm n}-\Sigma_{\rm an}-\mu}
  \label{7.3}
  \ea
  \ee
  may be considered as variational parameters instead
  of $\Sigma_{\rm n}$ , $\Sigma_{\rm an}$.

  The parameter $\delta$ should be set $\delta=1$ at
  the end of the calculations. This perturbation
  scheme is known as the $\delta$-expansion
  method  \cite{ramos}.

4) After subtraction of  discontinuous and one-particle reducible diagrams, we obtain
   the free energy $\Omega$ as a function of $\rho_{0}, X_{1}$
  and $X_{2}$.

  The variational parameters $X_{1}$ and $X_{2}$  may be
  fixed by the requirements
  \be
  \ba
  \displaystyle{
  \dsfrac{\partial\Omega(X_{1},X_{2},\rho_{0})}
   {\partial X_{1}}=0},\nonumber \\
   \\
   \displaystyle{
   \dsfrac{\partial\Omega(X_{1},X_{2},\rho_{0})}
   {\partial X_{2}}=0}.
   \label{7.4}
   \ea
   \ee
The condensed density $\rho_{0}$ it is determined by
   stationary condition
  \be
  \frac{d\Omega}{d\rho_{0}}=\frac{\partial\Omega}{\partial X_{1}}
  \dsfrac{\partial X_{1}}{\partial\rho_{0}}+
  \frac{\partial\Omega}{\partial X_{2}}
  \frac{\partial X_{2}}{\partial\rho_{0}}+
  \frac{\partial\Omega}{\partial\rho_{0}}=
  \frac{\partial\Omega}{\partial\rho_{0}}=0,
  \label{7.5}
  \ee
  that is, by partially differentiating $\Omega$ with respect
  to $\rho_0$ and setting it to zero.

  Note that (\ref{7.5}) is equivalent to the condition $\langle
  \widetilde{\psi}\rangle=0$, which is obtained by the requirement
  $H^{(1)}(\widetilde{\psi}, \widetilde{\psi}^{\dag})=0$ in the
  Hamiltonian formalism \cite{stoofbook}, where $H^{(1)}$ the part of
  the Hamiltonian which is linear at $\widetilde{\psi}$.

The accuracy of the $\delta-$expansion to calculate
$\Omega$ is somewhat limited by the fact that the inclusion of loop
integrals which are complicating the calculation process \cite{klbook,
stevenson}, is not carried out here.

\section*{Appendix B}
\def\theequation{B.\arabic{equation}}
\setcounter{equation}{0}

Here we present explicit expressions for ${E}_{k,T}^{\prime}
=d{E}_{k}/dT$ and
${E}_{k,\mu}^{\prime}=d{E}_{k}/d\mu$, which are
needed for the evaluation of the entropy and heat capacity in Equations (\ref{9.4})-(\ref{9.6}).
  In the normal phase when ${E}_{k}=\omega_{k}=\varepsilon_{k}
  -\mu+2U\rho$, the density of particles is given by
  \be
   \rho=\sum_{k}f_{B}(\omega_{k})\label{a1}
     \ee
  where $f_{B}(x)=1/(e^{\beta x}-1)$.
  Clearly,
  \be
  \frac{d\omega_{k}}{dT}=2U\frac{d\rho}{dT}\label{a2}
   \ee
   which does  not depend   on momentum $k$.
Differentiating both sides of the equation (\ref{a1}) with
  respect to $T$ and solving by $dp/dT$, we find
\bea
 \frac{d\rho}{dT}&=&\frac{\beta S_{1}}{2S_{2}-1},\nonumber\\
  S_{1}&=&-\beta \sum_{k}\omega_{k}f_{B}^{2}(\omega_{k})
   e^{\omega_{k}\beta},\nonumber\\
   S_{2}&=&-U\beta\sum_{k}f_{B}^{2}(\omega_{k})
   e^{\omega_{k}\beta}.\label{a3}
  \eea
Taking the derivative with respect to $\mu$ gives
  \bea
  \frac{d\omega_{k}}{d\mu}=2U\frac{d\rho}{d\mu}-1,\nonumber\\
   \frac{d\rho}{d\mu}=\frac{S_{2}}{U(2S_{2}-1)}.\label{a4}
    \eea
In the condensed phase, $T<T_{c}$, $E_{k}=\sqrt{\varepsilon_{k}(\varepsilon_{k}+2\Delta)}$, and hence we
  have
  \bea
  \frac{dE_{k}}{dT}=\frac{\varepsilon_{k}}{E_{k}}
   \Delta_{T}^{\prime},\nonumber\\
   \frac{dE_{k}}{d\mu}=\frac{\varepsilon_{k}}{E_{k}}
   \Delta_{\mu}^{\prime}.\label{a5}
   \eea

   To find, e.g., $\Delta_{T}^{\prime}$ we can differentiate
   both sides of the equation (\ref{11.5}) with respect to
   $T$ and solve it with respect to $\Delta_{T}^{\prime}$.

   The results are
   \bea
    &&  \Delta_{T}^{\prime}=\frac{d\Delta}{dT}=\frac{US_4}
   {2T(2S_{5}+1)},\nonumber \\
   && \Delta_{\mu}^{\prime}=\frac{d\Delta}{d\mu}=\frac{1}
   {2S_{5}+1},\nonumber \\
   && S_4=\sum_{k}W_{k}^{\prime}(\varepsilon_{k}+2\Delta),\nonumber\\
&& S_{5}=U\sum_{k}\frac{4W_{k}+E_{k}W_{k}^{\prime}}{4E_{k}},\\
&&  W_{k}^{\prime}=\beta(1-4W_{k}^{2}),\nonumber\\
   && W_{k}=\frac{1}{2}+f_{B}(E_{k}). \nonumber
\lab{bdelta}
\eea

In the HFP approximation, $\Delta^{\prime}_T$ is formally given by
\re{bdelta}, but with the following $S_4 $ and $S_5$:
\bea
 S_4\vert_{\rm HFP}&=&\sum_{k}W_{k}^{\prime}(\varepsilon_{k}+\Delta),\nonumber\\
 S_5\vert_{\rm HFP}&=&\frac{U}{4}\ds\sum_k\dsfrac{\veps_k(4\Delta W_k+E_k W^{\prime}_k(\Delta+\veps))}{E_{k}^{3}}.
\lab{deltashthfp}
\eea

Below we illustrate the low-temperature expansion explicitly. For this purpose we follow the strategy
outlined in Sect. III. and start with $\rho_1$. The Eq. \re{11.7} may be rewritten as
\be
\rho_1=\sum_{k}\frac{\veps_q+\Delta}{E_q(\exp(E_q\beta)-1)}+\rho_1(0),
\lab{rho11}
\ee
and its $T$ dependent part as
\be
I_1=\sum_{k}\frac{\veps_q+\Delta}{E_q(\exp(E_q\beta)-1)}=\frac{1}{4mc}\int_{0}^{Q_0}dq
\frac{q(q^2\pi^2+2m^2c^2)}{\exp({\pi cq\beta})-1}.
\lab{rho12}
\ee
This integral can be evaluated explicitly,
\bea
 I_1&=&\dsfrac{
\,T{Q_{0}}\, \left( {\pi }^{2}{{Q_{0}}}^{2}+2\,{m}^{2}{c}^{2}
 \right) \ln  \left( 1-{z}^{-1} \right)
}
{
4{m}{c}^{2}{\pi }
} +
{\dsfrac {
{T}^{2} \left( 3\,{\pi }^{2}{{Q_{0}}}^{2}+2\,{m}^{2}{c}
^{2} \right) {{\rm Li}_2} \left( {z}^{-1} \right) }
{
4m{c}^{3}{\pi }^{2}}
}+\nonumber\\
&&{\dsfrac {3{T}^{3} \left( -{Q_{0}}\,{{\rm Li}_3} \left( {z}^{-1}
 \right) c\pi +T{{\rm Li}_4} \left( {z}^{-1} \right)  \right) }{2m{c}^{5
}{\pi }^{2}}}-
\,{\dsfrac {{T}^{2} \left( {\pi }^{2}{T}^{2}+5\,{m}^{2}{c}^{4} \right) }{60m{c}^{5}}}-\\
&& {\dsfrac {{{Q_{0}}}^{2
} \left( {\pi }^{2}{{Q_{0}}}^{2}+4\,{m}^{2}{c}^{2} \right) }{16mc}},\nonumber
\lab{rho13}
\eea
where $z=\exp(-Q_0c\pi/T)$ and ${\rm Li}_s(z)=\ds\sum_{n=1}^{\infty}z^n/n^s$ is a polylogarithmic
function. Since $z\leq 1$ at  small $T$, we can perform an expansion
in $z$ and obtain
\be
I_1=\frac {m{T}^{2}}{12c}+
\frac {{\pi }^{2}{T}^{4}}{60m{c}^{5}}
- \left[
\frac{3{T}^{4}}{2m{c}^{5}{\pi}^{2}}+
\frac{3{T}^{3}Q_0}{2m{c}^{4}{\pi}}+
\frac{3{T}^{2}Q_{0}^{2}}{4m{c}^{3}}+
\frac{m{T}^{2}}{2{c}{\pi^2}}+
\frac{\pi TQ_{0}^{3}}{4m{c}^{2}}+
\frac{mT Q_{0}}{2\pi}
\right]z+O(z^2).
\lab{rho14}
\ee
The leading terms of this expansion are
\be
\rho_1=\rho_1(0)+\frac{\widetilde{T}^2}{12\gamma}+\frac{\pi^2\widetilde{T}^4}{60\gamma^5}+O(\widetilde{T}^6),
\lab{rho15}
\ee
where $\gamma=cm$. Using the expansion of the total magnetization  Eq. \re{Msm}, one may find
the low temperature expansion
for the total triplon density as
\be
 \rho=\rho(T)-\frac{\alpha_1}{4U\gamma m}\widetilde{T}^2-\frac{\alpha_3}{8Um\gamma^3}\widetilde{T}^4+O(\widetilde{T}^6).
\lab{rhotiot}
\ee
The previous two equations yield for the condensed fraction
\be
\rho_0=\rho-\rho_1=
 \rho_0(0)-\frac{(3\alpha_1+Um)}{12Um\gamma}\widetilde{T}^2-
 \frac{15\gamma c \alpha_3+2U\pi^2}{120U\gamma^5}\widetilde{T}^4
 +O(\widetilde{T}^6).
\lab{rho0}
\ee
Finally, excluding $\Delta$ from equations \re{rhotrho0}
gives the low temperature expansion for the anomalous density
\be
\sigma=\sigma(0)+\frac{Um-3\alpha_1}{12Um\gamma}\widetilde{T}^2+
\frac{2\pi^2U-15c\gamma\alpha_3}{120U\gamma^5}\widetilde{T}^4+O(\widetilde{T}^6).
\lab{sigmalow}
\ee
Low-temperature expansions for other quantities can be obtained in a similar way.

\section*{Appendix C  }
\def\theequation{C.\arabic{equation}}
\setcounter{equation}{0}

Here we will show that if
$r=(H-H_c)/H_c=\mu/\Delta_{\rm st}$ is small (where $\Delta_{\rm st}=g\mu_B H_c$
is the spin gap), the Gr{\"u}neisen parameter diverges as $\Gamma_H\sim 1/{r}$ at low temperatures.

To do this we study the second term of Eq. \re{ghsmex} which can be
written as
\be
\gamma_0=\dsfrac{2g\mu_B(3UQ_{0}^{2}\pi^2+12c\pi^2+10U\gamma^2-20U\gamma      ) }
{3\gamma^2 \pi^2(UQ_{0}^{2}+4c    )^2},
\lab{gr1}
\ee
where  $c^2=\Delta(0)/m$. The $\Delta(0)$ is given by Eq.
\re{11.5}, where $\sigma$ and $\rho_1$ are taken from \re{11.6} and  \re{11.7} with $W_k=1/2$,
and it can be simplified as
\be
\Delta=\mu+U\sum_{k}\left(1-\frac{E_k}{\veps_k}\right)
\lab{del02}
\ee
with $E_k=\sqrt{\veps_k}\sqrt{\veps_k+2\Delta}$. In the Debye-like approximation, the momentum integration in
\re{del02} can be taken explicitly, even without linear approximation for $E_k$, resulting in
\be
\Delta=\mu+\frac{U}{6\pi^2}
\left[
8(\Delta m)^{3/2}+\pi^3Q_{0}^{3}  - (\pi^2Q_{0}^{2}+4m\Delta )^{3/2}
\right].
\lab{del03}
\ee
It is clear that for small $r$, $\Delta$ becomes also arbitrarily
small, and $\pi^2Q_{0}^{2}+4m\Delta\approx \pi^2Q_{0}^{2}\approx 15.2$.
Thus the equation \re{del03} can be simplified to
 \be
 \Delta=\mu+\frac{4U(\Delta m)^{3/2}}{3\pi^2},
\lab{del04}
\ee
or in terms of $r$, to
\be
 \Delta=r\Delta_{\rm st}+\frac{4U(\Delta m)^{3/2}}{3\pi^2}.
\lab{del05}
\ee
For small $r$, the solution of this equation can be found by iteration,
\be
 \Delta\approx r\Delta_{\rm st}+r^{3/2}d_{32},
\lab{del06}
\ee
where $d_{32}=4U(m\Delta_{\rm st})^{3/2}/3\pi^2$ is  constant. Thus we come to the conclusion that
the velocity $c$ of the first sound is given by
$c=\sqrt{r}\sqrt{\Delta_{\rm st}/m}+O(r^{3/2})$. Inserting this into \re{gr1}
and taking only the leading term,
we obtain
 \be
\gamma_0=\frac{2}{UmQ_{0}^{2}(H-H_c)}+O(r).
\lab{gr11}
\ee

Thus, at low temperatures near $H_c$, the Gr{\"u}neisen parameter
scales as
\be
\Gamma_H\approx \frac{2}{UmQ_{0}^{2}(H-H_c)}.
\lab{grunsmal}
\ee

\bb{99}
\bi{wolf} B. Wolf \textit{et al.} Int. Journ. Mod. Phys. B {\bf 28}, 1430017 (2014).
\bi{Zhu} L. Zhu, M. Garst, A. Rosch, and Q. Si, Phys. Rev. Lett. {\bf 91}, 066404 (2003).
\bi{garst} M. Garst and A. Rosch, Phys. Rev. B {\bf 72}, 205129 (2005).
\bi{gegenw1} Y. Tokiwa and P. Gegenwart, Rev.  Sci. Instr. {\bf 82}, 013905 (2011).
\bi{gegenw2} P. Gegenwart arxiv: 1609.02013.
\bi{gegenrev} P. Gegenwart Rep. Progr. Phys. {\bf 79}, 114502 (2016).
\bi{zapf} V. Zapf , M. Jaime and C. D. Batista Rev. Mod. Phys. {\bf 86}, 563 (2014).
\bi{giam} T. Giamarchi, C. Ruegg, and O. Tchernyshyov, Nature Physics {\bf 4}, 198 (2008).
\bi{dtn2012} F. Weickert, \textit{et al.}, Phys. Rev. B {\bf 85}, 184408 (2012).
\bi{zvyagin}  S. A. Zvyagin,   Phys. Rev. Lett. {\bf 98}, 047205 (2007).
\bi{nikuni} T. Nikuni, M. Oshikawa, A. Oosawa, H. Tanaka, Phys. Rev. Lett. {\bf 84} , 5868 (2000).
\bi{ourannals}  Rakhimov A. , Mardonov S. , and Sherman E. Ya.  Ann. Phys. {\bf 326}, 2499 (2011).
\bi{ourprb} Rakhimov A. , Sherman E. Ya., and Kim Chul Koo  Phys. Rev. B {\bf 81}, 020407(R) 2010.
\bi{ournjp1}Rakhimov A. \textit{et al.} New J. Phys. {\bf 14}, 113010 (2012).
\bi{ournjp2}  A. Khudoyberdiev , A. Rakhimov and A.
Schilling, New J. Phys. {\bf 19}, 113002 (2017).
\bi{andersen} J. Andersen   Rev. Mod. Phys. {\bf 76}, 599 (2004).
\bi{matsubara} T. Matsubara and H. Matsuda, Prog. Theor. Phys. {\bf 16},
 569 (1956).
\bi{matsumoto}  M. Matsumoto \textit{et al.}, Phys. Rev. B {\bf 69}, 054423 (2004).
\bi{ourlatt}H. Kleinert, Z. Narzikulov, and A. Rakhimov,
Phys. Rev. A {\bf 85}, 063602 (2012).
\bi{bellac} M. Le Bellac, Thermal
Field Theory (Cambridge University Press, Cambridge, 1996).
\bi{klbook} H. Kleinert and V. Schulte-Frohlinde, Critical Phenomena in
$\phi^4$-Theory (World Scientific, Singapore, 2001).
\bi{ourjstatmeh} H. Kleinert, Z. Narzikulov and A. Rakhimov,
 J. Stat. Mech. P01003 (2014).
\bi{ourpra77} A. Rakhimov, Chul Koo Kim, Sang-Hoon Kim, and Jae Hyung Yee
Phys. Rev. A {\bf 77}, 033626 (2008).
\bi{yamada} F. Yamada   et al., J. Phys. Soc. Japan {\bf 77},  013701  (2008).
\bi{pines} N. M. Hugenholtz  and D.  Pines  Phys. Rev. {\bf 116}, 489 (1959).
\bi{redbook}T. Haugset, H. Haugerud, and F. Ravndal, Ann. Phys. {\bf 266}, 27
(1998).
\bi{misgu} G.  Misguich  and M. J.  Oshikawa  Phys. Soc. Jpn. {\bf 73} , 3429 (2004);\\
 R. DellAmore, A. Schilling  and K. Kramer   Phys. Rev. B {\bf 79}, 014438 (2009).
\bi{yuklaser} V. I. Yukalov, Laser Physics {\bf 19}, 1 (2009), (See chapter 7).
\bi{robinson} J. E. Robinson Phys. Rev, {\bf 83}, 678 (1951).
\bi{helium} A. Fetter and J. Walecka Quantum theory of many particle system (Dover Publications, NY, 2003).
\bi{ouriman} A. Rakhimov and I. N. Askerzade, Int. Journ.  Mod. Phys.
 B  {\bf 29},  1550123   (2015);
 \\ A. Rakhimov and I. N. Askerzade, Phys. Rev. E {\bf 90}, 032124 (2014).
\bi{huangbook} K. Huang Statistical Physics (John Wiley \& Sons, 1997).

\bi{prokofev} Z. Yao \textit{et al.} Phys. Rev. Lett. {\bf 112}, 225301, (2014).
\bi{aczelSr} A. A. Aczel \textit{et al.} Phys. Rev. Lett. {\bf 103}, 207203 (2009).
\bi{kofuprl} M. Kofu \textit{et al.} Phys. Rev. Lett. {\bf 102}, 177204 (2009).
\bi{wangprl}Zhe Wang \textit{et al.} Phys. Rev. Lett. {\bf 116}, 147201 (2016).
\bi{tanaka} H. Tanaka \textit{et al.}, J. Magn. Mag. Mater. {\bf 310},  1343 (2007).
\bi{kohamamce} Y. Kohama \textit{et al.} Rev. Scientific Instr. {\bf 81}, 104902 (2010).
\bi{schill} A. Schilling and M. Reibelt, Rev. Sci. Instrum. {\bf 78},
033904   (2007).
\bi{wangprb} Z.  Wang Phys. Rev. B {\bf 85}, 224304, (2012).
\bi{ramos} Frederico F. de Souza Cruz et al. Phys. Rev.  B{\bf 64}, 014515 (2001).
\bi{stoofbook} H. T. C. Stoof, K. B. Gubbels and D.B.M. Dickerscheid Ultracold Quantum Fields (Springer, 2009).
\bi{stevenson} I. Stancu and P. M. Stevenson Phys. Rev. D{\bf 42}, 2710 (1990).
\eb
\edc